\documentclass[prd,nofootinbib,floats,superscriptaddress,eqsecnum,tightenlines,preprintnumbers,11pt]{revtex4}

\usepackage{hyperref}
\usepackage{graphicx}
\usepackage{amsmath,amssymb,amsfonts,amsthm,latexsym}
\usepackage{marginnote}
\usepackage{xcolor}
\usepackage{physics}
\usepackage{nicefrac}

\usepackage{amsmath}%
\usepackage{amsfonts}%
\usepackage{amssymb}%
\usepackage{graphicx}

\usepackage{mathrsfs}
\usepackage{xspace}
\usepackage{mathtools, amssymb}

\usepackage{tikz}
\tikzset{every picture/.style={line width=0.75pt}} %set default line width to 0.75pt   

\def\be{\begin{equation}}
\def\ee{\end{equation}}
\def\beq{\begin{equation}}
\def\eeq{\end{equation}}
\newcommand{\bea}{\begin{eqnarray}}
\newcommand{\eea}{\end{eqnarray}}
\def\bi{\begin{itemize}}
\def\ei{\end{itemize}}
\def\ba{\begin{array}}
\def\ea{\end{array}}
\def\bfig{\begin{figure}}
\def\efig{\end{figure}}

\newcommand{\A}{{\cal A}}   % A-graphics: 3 links shared by the vertices
\newcommand{\Abis}{\tilde{\cal A}}   
\newcommand{\At}{\tilde{\cal A}}   
\newcommand{\B}{{\cal B}}   % B-graphics: 1 link shared by the vertices
\newcommand{\Bbis}{\tilde{\cal B}}   
\newcommand{\Bter}{\hat{\cal B}}   
\newcommand\f{f_2} % Short notation for <F_2>
\newcommand\g{g_4} % Short notation for coefficient of g in Cayley-Hamilton

\newcommand{\F}{F}  % matrix F

\newcommand{\R}{\mathcal{R}}  % Riemann terms
\newcommand{\U}{\tilde V}
\newcommand{\V}{V}
\newcommand{\W}{W}

\newcommand{\AR}{A}
\newcommand{\BFF}{B}
\newcommand{\s}{\lambda}

\begin{document}
\title{Classification of generalised higher-order Einstein-Maxwell Lagrangians}

\author{Aimeric Coll\'{e}aux}
\affiliation{Astrophysics Research Center, The Open University of Israel, Ra’anana, Israel}
\author{David Langlois}
\affiliation{Universit\'e Paris Cit\'e, CNRS, Astroparticule et Cosmologie, F-75013 Paris, France}
\author{Karim Noui}
\affiliation{Laboratoire de Physique des deux Infinis IJCLab, Universit\'e Paris Saclay, CNRS, France}

\date{\today}

\begin{abstract}
We classify all higher-order generalised Einstein-Maxwell Lagrangians that include terms linear in the curvature tensor and quadratic in the derivatives of the electromagnetic field strength tensor. Using redundancies due to the Bianchi identities, dimensionally dependent identities and boundary terms, we show that a general Lagrangian of this form can always be reduced to a linear combination of only 21 terms, with coefficients that are arbitrary functions of the two scalar invariants derived from the field strength. We give an explicit choice of basis where these 21 terms include  3 terms linear in the Riemann tensor and 18 terms quadratic in the derivatives of the field strength. 
\end{abstract}

\maketitle

%\tableofcontents

\section{Introduction}
The construction of scalar-tensor theories and vector-tensor theories with non-minimal coupling to the spacetime curvature in 4 dimensions was explored systematically by Horndeski almost half a century ago \cite{Horndeski:1974wa,Horndeski:1976gi}.  In doing so, he imposed the apparently reasonable  restriction that the equations of motion should be second-order. 
This condition was believed, until a few years ago,  to be  indispensable to avoid the presence of  extra degrees of freedom, known as Ostrogradski ghosts, which are expected to lead to problematic instabilities\footnote{From the EFT (Effective Field Theory) point of view, extra degrees of freedom of this type  might be acceptable, provided  they cannot be excited in the regime of validity of the theory.}.

However, it was later realised that imposing second-order equations of motion turns out to be too restrictive. Indeed, as shown explicitly in the context of scalar-tensor theories, it is possible to construct Lagrangians that lead to higher-order (third or fourth order) equations of motion but do not contain any extra degree of freedom, provided they satisfy some degeneracy conditions~\cite{Langlois:2015cwa,Langlois:2015skt}\footnote{See  \cite{Gleyzes:2014dya,Gleyzes:2014qga} for earlier works pointing to viable theories beyond the Horndeski family, although it was not yet realised that degeneracy was the key ingredient to avoid extra degrees of freedom.}. These theories, dubbed DHOST theories in \cite{BenAchour:2016cay}, were fully classified up to quadratic order (in second derivatives of the scalar field) in  \cite{Langlois:2015cwa} and up to cubic order  later in \cite{BenAchour:2016fzp} (see \cite{Langlois:2018dxi} and  \cite{Kobayashi:2019hrl} for reviews).  This type of construction was  also considered in  the case of vector-tensor theories of gravity, dubbed generalised Proca theories  \cite{Tasinato:2014eka,Heisenberg:2014rta,Allys:2015sht,BeltranJimenez:2016rff}, as well as 
in higher order extensions of Einstein-Maxwell theory from disformal transformations  \cite{Gumrukcuoglu:2019ebp,DeFelice:2019hxb}.
In the present work, we focus our attention on $U(1)$ vector-tensor theories non-minimally coupled to gravity, including derivatives of the field strength in their Lagrangian.

From the results of Horndeski in \cite{Horndeski:1976gi}, it can be seen that in flat four-dimensional spacetime, there is no theory quadratic in the derivatives of the field strength which yields second order field equations. Note that this result has been partially extended to arbitrary dimensions in \cite{Deffayet:2013tca}. Historically, the first theory of this type has been proposed many decades ago by Bopp and Podolsky in \cite{Bopp1,Podolsky:1942zz} as an improvement of the UV behaviour of classical electrodynamics. For instance, the Coulomb potential is non-singular at the origin in this model. 
More general Lagrangians quadratic in the derivative of the field strength can also be found in the context of quantum electrodynamics (QED), where the derivative expansion of the one-loop effective action, obtained by integrating out the electron in an external slowly varying electromagnetic field precisely generate such effective self-interactions,  generalising the famous Euler-Heisenberg action \cite{Lee:1989vh,Cangemi:1994by, Gusynin:1995bc,Gusynin:1998bt,Navarro-Salas:2020oew,Karbstein:2021obd}. 

As far as effective interactions between gravity and electromagnetism are concerned, the one-(electron)-loop effective action in the presence of weak external gravitational and electromagnetic fields has been found by Drummond and Hathrel \cite{Drummond:1979pp}. Its generalisations for strong external and slowly varying fields have been investigated in \cite{Bastianelli:2007me,Bastianelli:2008cu} and precisely fall into the family of Lagrangians we are going to study. Interesting implications of these QED-induced non-minimal couplings are ranging from superluminality  \cite{deRham:2020zyh} to gravitational birefringence and their effect on gravitational waves \cite{Ejlli:2018hke, Ejlli:2020fpt}. New non-minimally coupled theories, which preserve the duality-invariance of Maxwell theory and conformal electrodynamics (see \cite{Bandos:2020jsw,Kosyakov:2020wxv}), have also been obtained recently \cite{Cano:2021tfs,Cano:2021hje}. Interestingly, much like non-minimal couplings involving a non-abelian gauge field \cite{Balakin:2015gpq}, some models yield non-singular black hole solutions \cite{Cano:2020ezi}.

 In the present article, our goal is to  classify  all possible Lagrangians that are linear in the curvature and quadratic in derivatives of the field strength tensor, without investigating at this stage the presence and nature of potential extra degrees of freedom. 
 A first exploration of the presence of extra degrees of freedom will be considered for the flat case in a future work.
Even limiting the Lagrangians to terms linear in the curvature and quadratic in the field strength derivatives, the number of possible terms appears enormous. Fortunately, there exists a number of redundancies, which we exploit systematically in order to drastically reduce the number of terms that need to be taken into account in a generic Lagrangian. 
These redundancies follow from a) the index symmetries of the various possible scalar terms; b) the Bianchi identities resulting from the $U(1)$ symmetry of the vector field; c) the Bianchi identities of the Riemann tensor ; d) the dimensionally dependent identities (DDIs); e) the boundary terms. 

Taking into account the first four redundancies, which are purely algebraic, we show that the most general Lagrangian of our interest can be written as a linear combination of 4 Riemann-linear terms and 22 field-strength-quadratic terms,  whose coefficients are arbitrary functions of the two scalar invariants constructed from the field-strength tensor. We finally find  five independent  boundary terms  of the appropriate form, which enables us  to further reduce the number of terms in the minimal Lagrangian to 21.

\medskip
The structure of this article is the following. In the next section, we present the general action describing the theories we are interested in, introduce some useful notations, and explain how we will proceed to eliminate redundant terms. In section \ref{section_Riemann}, we examine the Riemann-dependent terms and show that they can be reduced to only 4 terms. We then turn, in section \ref{section_DFDF} to the major part of this work, where we show that all the terms quadratic in the derivatives of the field strength can be represented by only 22 terms. After this purely algebraic analysis, we discuss, in section \ref{section_boundary}, how boundary terms can also reduce the minimal number of terms in the most general Lagrangian: we identify 5 relevant boundary terms. As summarised in section \ref{section_basis}, this means that the most general Lagrangian can be written as a linear combination of only 21 terms, up to arbitrary coefficients that are scalar functions of the two electromagnetic invariants. We conclude in the final section. More technical aspects on our derivations are provided in an appendix.

\section{Higher-order Einstein-Maxwell theories}

We consider the most general family of four-dimensional abelian vector-tensor theories that are linear in the curvature tensor\footnote{Our convention for the Riemann tensor is such that $(\nabla_a\nabla_b-\nabla_b\nabla_a)v_c=R_{abc}{}^dv_d$.} 
and quadratic in the covariant derivative of the Faraday tensor, or field strength,
\beq
\label{F}
F_{ab}=\nabla_aA_b-\nabla_b A_a
\eeq
 where $A_a$ is the gauge potential vector. The action describing these theories can be written in the form
\begin{eqnarray}
\label{action}
S= \int d^4 x \sqrt{-g} \left(L_0 + \frac{1}{4} \AR^{abcd} R_{abcd} + \BFF^{abc,def} \nabla_a F_{bc}\nabla_d F_{ef}  \right), \label{Action}
\end{eqnarray} 
where the scalar ${L}_0$ and the tensors $\AR^{abcd}$ and $\BFF^{abc,def}$  contain all possible terms  that can be built from combinations of  the metric $g_{ab}$, of the
Levi-Civita tensor $\varepsilon_{abcd}$ and of  the Faraday tensor $F_{ab}$. Of course, the tensors $\AR^{abcd}$ and $\BFF^{abc,def}$ can be simplified by taking into account the symmetries of the Riemann tensor or of the term quadratic in the covariant derivative of $F_{ab}$: for instance, any term in $\AR^{abcd}$ that is symmetric with respect to the indices $a$ and $b$ is irrelevant since it will disappear after contraction with the Riemann tensor.

Let us first recall that one can construct two scalars from the field strength tensor (\ref{F}), which correspond to the two well-known   Lorentz invariants available in standard electromagnetism, namely 
\begin{eqnarray}
\f \equiv F^{ab}F_{ab} \,, \qquad f_*\equiv {}^*\!F_{ab}\,  F^{ab},
\end{eqnarray}
where 
\beq
{}^*\!F_{ab} \equiv  \frac{1}{2}\varepsilon_{abcd} F^{cd}
\eeq
 is the dual field strength.

 In order to use compact notations, it will be very useful to define the tensors $F_{I}^{ab}$ where the index $I$ corresponds to the number of contracted Faraday tensors, i.e.
\begin{eqnarray}
F^{a}_{2\, b} \equiv F^{a}_{\ c} F^{c}_{\ b}\,, \qquad  F^{a}_{3\, b} \equiv F^{a}_{\ c} F^{c}_{\ d} F^{d}_{\ b}\,, \quad  {\rm etc}\, ,
\end{eqnarray}
and, for later convenience, we also generalise this notation to include the metric and the field strength, so that
\beq
F^{a}_{0\, b} \equiv g^{a}_{\ b}\,, \qquad  F^{a}_{1\, b} \equiv F^{a}_{\ \ b}\,.
\eeq

It is worth noting that, due to the Cayley-Hamilton relation in four dimensions, any $F_K^{ab}$ with  $K\geq 4$ can always be written in terms of the first four tensors of the tower, namely $F_I^{ab}$ with $0\leq I\leq 3$. Indeed,
the identity 
 \begin{eqnarray}
\delta^a_{[b} F^{e_1}{}_{e_1} F^{e_2}{}_{e_2} F^{e_3}{}_{e_3} F^{e_4}{}_{e_4]} =0\,, 
\end{eqnarray}
where the brackets denote antisymmetrisation over the corresponding indices,
leads to  the Cayley-Hamilton relation
\beq
\label{CH}
F^{a}_{4\, b}+ \frac12 \f F^{a}_{2\, b} -\frac{1}{4}  (f_4 - \frac{1}{2} \f^2 )  \delta^a_b=0\,,
\eeq
where  we have introduced 
\beq
f_4\equiv F_2^{ab}F_{2 ab}=F^a_{4a}\,.
\eeq
This implies that  $F^{ab}_{4}$ can be rewritten as a linear combination of  lower weight\footnote{Throughout this paper, the weight of any scalar appearing in the action \eqref{Action} denotes the number of field strength $F$ contained in $\AR$ or $\BFF$.}  tensors $F^{ab}_2$ and $g_{ab}$, with coefficients that depend on the two  scalars $\f$ and $f_4$. 

By  induction, all tensors $F^{ab}_{I}$  with $I$ even can similarly be expressed  as linear combinations of the two tensors $F^{ab}_2$ and $g_{ab}$ with coefficients that are functions of the two scalars $\f$ and $f_4$. Moreover, by contracting \eqref{CH} by $F^b_{\ c}$, one finds that $F^{ab}_{5}$ can be written in terms of $F^{ab}_3$ and  $F^{ab}$ and of the scalars $\f$ and $f_4$. Again by induction, the same applies to any tensor $F^{ab}_I$ with $I$ odd.

\medskip
As mentioned earlier, we allow the Lagrangian terms in (\ref{action}) to include a  dependence on the Levi-Civita tensor $\varepsilon_{abcd}$. Since 
\beq
\varepsilon_{abcd}\, \varepsilon^{efgh} = -\delta^{efgh}_{abcd}\,,
\eeq
where $\delta$ on the right hand side is the generalised Kronecker symbol (which corresponds to an antisymmetrised product of usual Kronecker symbols),
 any tensor that is constructed with an even number of $\varepsilon_{abcd}$, which we call an even-parity tensor, can be equivalently
expressed without resorting to $\varepsilon_{abcd}$. For odd-parity tensors,  i.e. which contains an odd number of  $\varepsilon$, they can be rendered even by multiplication with any odd scalar, the simplest being $f_*$ introduced earlier.  For example, multiplying $f_*$ by itself, one gets the even scalar 
 \beq
 f_*^2=4 \left(f_4-\frac12 \f^2\right) \equiv 4 g_4\,.
 \eeq
As a consequence, we can safely ignore any dependence on $\varepsilon_{abcd}$ in the rest of our analysis \footnote{Indeed, given an odd scalar $S=\varepsilon_{abcd} S^{abcd}$, where the tensor $S^{abcd}$ is  of even parity, and a generic function $\Phi$ of $f_2$ and $f_4$, we can write 
 \begin{eqnarray}
 \Phi S = \widetilde{\Phi} \delta^{efgh}_{abcd}  F_{ef}F_{gh} S^{abcd} \equiv \widetilde{\Phi}\widetilde{S}
 \end{eqnarray}
where $\widetilde{S}$ is a parity even scalar and $\widetilde{\Phi}=\Phi/(4 \sqrt{g_4})$, choosing the negative branch of the square-root.}.

\medskip
From our discussion above, we conclude that  all possible Lagrangian terms can be written as  combinations of the form
\begin{eqnarray}
\R_{IJ} \equiv F^{ab}_I F^{cd}_J \left( \alpha_{IJ} R_{abcd} + \beta_{IJ} R_{acbd} \right) \;,\;\;\;\;\mathcal{F}_{IJK}  = \sum_{i} \gamma^{(i)}_{IJK} F^{ab}_I F^{cd}_J F^{ef}_K \left( \nabla F \nabla F \right)^{(i)}_{abcdef}  \,,\;\;\; 
\end{eqnarray}
where $I,J,K \in\{0,1,2,3\}$ and the coefficients $\alpha, \beta, \gamma$ are arbitrary scalar  functions of $\f$ and $f_4$. The last term on the right-hand side of the second relation denotes all possible combinations of  the six free indices in $\nabla \F\nabla \F$.

\medskip
Modulo the  arbitrary scalar coefficients, the number of terms of the above form is finite but 
still remains high. As we will see below, this number can be further reduced by systematically taking into account various identities. The first category of identities are the Bianchi identities, both for the  Riemann tensor,
\beq
\label{Bianchi1_R}
R_{[abc]d}=0\,,
\ee
and for the
field strength
\beq
\label{Bianchi_F}
\nabla_{[a}F_{bc]}=0\,.
\eeq

Other useful identities are the dimensionally dependent identities (DDIs) \cite{Edgar:2001vv}. These identities simply come from the fact that, in a 4-dimensional spacetime, all tensors that are antisymmetric with respect to 5 indices, or more, automatically vanish, so that 

\beq
\label{DDI_tensor}
\Lambda^{\dots}_{\dots[a_1a_2a_3a_4a_5]}=0\,,
\eeq
where the dots stand for possible additional indices, which are arbitrary. The Cayley-Hamilton identity (\ref{CH}) is one particular example. In this way, one can  construct {\it scalar} DDIs by contracting all indices, i.e. writing
\beq
\label{DDI_scalar}
\Lambda^{a_1a_2a_3a_4a_5}_{[a_1a_2a_3a_4a_5]}
=0\,,
\eeq
where $\Lambda$ is an arbitrary 10-index tensor.
Note that any quantity involving more than five antisymmetrised indices can be reduced to this form by Laplace's formula for determinants. Moreover, if the tensor $\Lambda$ contains a  Kronecker symbol, this  would produce an overall $(d-4)$ factor, making the corresponding identity trivial.

Finally, we will also use the possibility to relate different terms up to boundary terms, since the latter are irrelevant in the equations of motion.

\section{Riemann-dependent terms}
\label{section_Riemann}

We start by the classification of all the terms that are linear in the Riemann tensor, of the form
\beq
\label{R_terms_generic}
\R_{IJ} \equiv F^{ab}_I F^{cd}_J \left( \alpha_{IJ} R_{abcd} + \beta_{IJ} R_{acbd} \right)\equiv   \R^{(\alpha)}_{IJ}+\R^{(\beta)}_{IJ}\,.
\eeq
It can be checked that the sum of the indices $I+J$ has to be even, otherwise one of the indices, say $I$, is even and the other one is odd, which implies that $\R_{IJ}$ vanishes due the symmetries of the Riemann tensor since $\F_I$ is symmetric and $\F_J$ antisymmetric. Moreover the terms $\R^{(\alpha)}_{IJ}$ vanish when either $I$ or $J$ is even, again because of the symmetries of the Riemann tensor. The non-trivial terms that remain are thus $\R^{(\beta)}_{00}$, $\R^{(\beta)}_{02}$, $\R^{(\beta)}_{22}$, $\R_{11}$, $\R_{13}$ and $\R_{33}$. 

Taking into account the first Bianchi identities for the Riemann tensor \eqref{Bianchi1_R}, we obtain the relation
\beq
\label{Bianchi_FFR}
 F^{ab}_I F^{cd}_J \left( R_{abcd} -2 R_{acbd} \right)=0\,, \qquad  {\rm for} \ I \ {\rm and }\ J \ {\rm odd}
\eeq
which shows that all the $\R^{(\beta)}_{IJ}$ can be reexpressed as $\R^{(\alpha)}_{IJ}$ 
when both $I$ and $J$ are odd.
We are now left with   $\R^{(\beta)}_{00}, \R^{(\beta)}_{02}, \R^{(\beta)}_{22}, \R^{(\alpha)}_{11}, \R^{(\alpha)}_{13}$ and  $\R^{(\alpha)}_{33}$. 

Finally, we investigate whether some of these terms could be related by scalar DDIs of the form (\ref{DDI_scalar}), when the tensor $\Lambda$ contains the Riemann tensor and products of the field strength. Due to the Bianchi identities (\ref{Bianchi1_R}),  the antisymmetrisation  must involve at most  two indices of the Riemann tensor, with the other indices coming from the $\F_I$.  In order to reach a total of 10 indices, this means that the tensor $\Lambda$ must contain at least three $\F_I$, which leads us to relations of the form\footnote{Note that the terms $F^{a_1}_{I[a_1} F^{a_2}_{Ja_2}  F^{a_3}_{Ka_3}F^{a_4}_{La_4}R^{a_5}_{a_5]}$    can have the colour indices $(IJKL)=(1111)$ or $(2211)$. In both cases the result is trivial.}
\begin{eqnarray}
\label{DDI_Riemann}
\mathscr{R}_{IJK} \equiv 
F^{a_1}_{I[a_1} F^{a_2}_{Ja_2}  F^{a_3}_{Ka_3}  \left( \alpha_{IJK} R_{a_4a_5]}{}^{a_4a_5} + \beta_{IJK} R^{a_4}{}_{a_4a_5]}{}^{a_5}\right) =0\,,
\end{eqnarray}
where $I,J,K \in \{1,2,3\}$ (we do not include the index $0$ since it would induce a delta function and therefore a trivial identity) and $4 \leq I+J+K \leq 6$ (since the terms $\R_{LM}$ are at most of weight $6$, i.e. $L+M\leq 6$).  Even if three $\F_I$ appear in the above expressions, the expansion  of (\ref{DDI_Riemann}) and the contraction of all indices necessarily leads to final terms with at most two $F_I$ contracted with the Riemann tensor, i.e. exactly of the form (\ref{R_terms_generic}).

 Eventually we find only  two non-trivial expressions from (\ref{DDI_Riemann}). The first, corresponding to $(IJK)=(112)$, reads
\begin{eqnarray}
\label{DDI_Riemann_112}
F^{ab}_2 F^{cd}_2 R_{acbd} -2 F^{ab} F^{cd}_3 \left(R_{acbd}-R_{abcd}\right)+\frac12 \f F^{ab} F^{cd} \left(R_{acbd} - R_{abcd} \right)+\frac14 \g R=0
\end{eqnarray}
and can be further simplified using (\ref{Bianchi_FFR}) to yield
\beq
\label{DDI_Riemann_112bis}
2 F^{ab}_2 F^{cd}_2 R_{acbd} +2 F^{ab} F^{cd}_3 R_{abcd}-\frac12 \f F^{ab} F^{cd} R_{abcd} +\frac12\g R=0\,.
\eeq
The second relation, corresponding to $(IJK)=(123)$, yields, after simplification with (\ref{Bianchi_FFR})
\beq
\label{DDI_Riemann_123}
F^{ab}_3 F^{cd}_3 R_{abcd}+ \f F^{ab}_2 F^{cd}_2 R_{acbd}+\frac14 \g F^{ab} F^{cd} R_{abcd}+ \g F^{ab}_2 R_{ab}=0\,.
\eeq
The second relation tells us that the term $F^{ab}_3 F^{cd}_3 R_{abcd}$, of weight 6, can always be replaced by terms of lower weight, up to some scalar multiplicative coefficients\footnote{Of course, the total weight, when the weight of the scalar coefficients is taken account, is unchanged.}, and therefore does not have to be considered in the Lagrangian. Similarly, the  relation (\ref{DDI_Riemann_112bis}) tells us that one of the first two terms  can be expressed in terms of the other one and of  lower weight terms.

In summary, all the possible  Riemann-linear terms in the action can be written as linear combinations of only four terms, with scalar coefficients that depend on $\f$ and $f_4$, 
\begin{eqnarray}
\label{basis_Riemann}
\mathcal{R}_1\equiv R  \;,\;\;\;\;
\mathcal{R}_2\equiv F^{ab} F^{cd} R_{abcd} \;,\;\;\;\;\mathcal{R}_3\equiv F_{2}^{ab} R_{ab}  \;,\;\;\;\;\mathcal{R}_4\equiv  F^{ab}F_3^{cd} R_{abcd} \;\; ({\rm or}\;\; F_{2}^{ab}F_{2}^{cd} R_{acbd})  \,.
\end{eqnarray}
We have kept an alternative choice for the last term, but we will see later that this term can be eliminated in favour of other terms in the Lagrangian, thanks to boundary terms.

\section{Quadratic terms in $\nabla F$}
\label{section_DFDF}

The goal of this section is to determine the minimal number of terms of the form 
\beq
\BFF^{abc,def} \nabla_a F_{bc}\nabla_d F_{ef} \,,
\label{DFDF}
\eeq
taking into account the index symmetries, the Bianchi identities and  the dimensional identities. 

\medskip
For  a picturial representation of these terms, one can imagine that each $\nabla_e F_{ab}$ is a three-limb vertex, with a single leg corresponding to the covariant derivative index $e$ and  two arms associated with the indices of the antisymmetric tensor  $F_{ab}$.  Each term in \eqref{DFDF} thus corresponds to  two vertices whose limbs  
 are all connected, and  ``dressed", by  ``coloured" links that consist of
\beq
F_I^{ab} \quad \text{with}\qquad I \in \{ 0,1,2 ,3\} \, .
\eeq
In the following,  we will call the indices $I$ ``colour" indices to distinguish them from the spacetime indices denoted by a lowercase latin letter.

\subsection{Two main families of terms}
\label{familiesAandB}
Among all possible  scalar terms, it is useful to distinguish the family where the two vertices share all three links, family which can be subdivided into two subfamilies:
\beq
\label{Agraph}
\A_{IJK}=F_I^{ad}F_J^{bf} F_K^{ec} \nabla_e F_{ab} \nabla_f F_{cd}
\eeq
and
\beq
 \Abis_{IJK}=F_I^{ad}F_J^{bc} F_K^{ef} \nabla_e F_{ab} \nabla_f F_{cd}\,.
\eeq
In the first subfamily,  each leg is connected to one of the arms  of the other vertex, whereas in the second subfamily, the two vertex legs are interconnected. 

It is immediate to check that the above terms satisfy the following symmetries under permutation of two colour indices:
\beq
\A_{IJK}=(-1)^{I+J+K}\A_{IKJ}\,, \qquad \Abis_{IJK}=\Abis_{JIK}=(-1)^{I+J+K}\Abis_{JIK}\,,
\eeq
which implies that $I+J+K$ must be even to get a nonzero $\Abis$.
As a consequence of these symmetries, one finds 32 independent $\A$ terms\footnote{The nonzero independent terms are described by e.g. $J\leq K$ for $I$ even, which gives $10\times 2=20$ terms and $J<K$ for $I$ odd, which gives $6\times 2=12$ terms. Hence a total of $32$.},  and 20 independent $\Abis$ terms\footnote{The nonzero independent $\Abis$ terms are described by  $I\leq J$ with $I+J$ even  for $K$ even, which gives $6\times 2=12$ terms, and $I<J$ with $I+J$ odd for $K$ odd, which gives $4\times 2=8$ terms. Hence a total of $20$.}.

For the other family of scalar terms, only one link is shared by the two vertices, which leads to three subfamilies, depending whether this link connects two arms, two legs or one leg with an arm, respectively. The corresponding definitions are given by
\beq
\label{Bgraph}
\B_{IJK}=F_I^{ad}F_J^{eb} F_K^{cf} \nabla_e F_{ab} \nabla_f F_{cd}
\eeq
and
\beq
\Bbis_{IJK}=F_I^{ef}F_J^{ab} F_K^{cd} \nabla_e F_{ab} \nabla_f F_{cd}\,, \qquad \Bter_{IJK}= F_I^{bf}F_J^{ae} F_K^{cd} \nabla_e F_{ab} \nabla_f F_{cd}\,.
\eeq
These new terms satisfy the following symmetries under exchange of the colour indices,
\beq
\label{symmetries_B}
\B_{IJK}=(-1)^{I+J+K}\B_{IKJ},, \qquad \Bbis_{IJK}=(-1)^I\Bbis_{IKJ}\,.
\eeq
Moreover, by exchanging the two arms of a vertex  whenever they are interconnected, one obtains the following identities, 
\beq
\Bbis_{IJK}=(-1)^{J+1}\Bbis_{IJK}=(-1)^{K+1}\Bbis_{IJK}\,,\qquad
\Bter_{IJK}=(-1)^{K+1}\Bbis_{IJK}\,,
\eeq
which simply state that the colour indices $J$ and $K$ must be odd in $\Bbis_{IJK}$, whereas $K$ must be odd in $\Bter_{IJK}$. As a consequence of these symmetries, we find 32 independent $\B$ terms (since they have the same symmetries as the $\A$ terms), 8 independent $\Bbis$ terms\footnote{For $I$ even, the independent $\Bbis$ terms are described by $J\leq K$ , with $J$ and $K$ odd, which gives $2\times 3=6$ terms. For $I$ odd, the independent $\Bbis$ terms are described by $J< K$ , with $J$ and $K$ odd, which gives $2$ terms. Hence a total of $8$.}   and 32 independent $\Bter$ terms\footnote{Since $K$ is odd, we have $4\times 4\times 2=32$ terms.}. It is easy to convince oneself that all possible terms can be rewritten into one of the five forms above, up to some trivial permutation of spacetime indices. Overall, given a two-form $F$ in four dimensions, there are $124$ scalars of the form \eqref{DFDF}.

\subsection{Bianchi identities}
Let us now exploit the Bianchi identities (\ref{Bianchi_F}) satisfied by the Faraday tensor, which can be written as
\beq
\nabla_e F_{ab} +\nabla_a F_{be} +\nabla_b F_{ea} =0\,.
\eeq
Replacing $\nabla_e F_{ab}$  in the definition  (\ref{Agraph}) of $\A$ by  the above relation, one  obtains the sum of three terms, which are deduced from each other by a ``rotation'', i.e. a circular permutation, of the limbs of the first vertex. 
After  some straightforward reshuffling of the indices, this leads to the relation
\beq
\A_{IJK}+\A_{KJI}-\Abis_{IKJ}=0\,.
\eeq
As a consequence, one finds that all the $\Abis$ terms can be reexpressed in terms of the $\A$ terms via the relation
\beq
\label{constraint_Abis_Bianchi}
\Abis_{IJK}=\A_{IKJ}+\A_{JKI}\,.
\eeq
Moreover, for the indices $(IJK)$ such that $I+J+K$ is odd, we have seen that $\Abis_{IJK}=0$.  One thus  obtains additional relations, which provide  10 independent  constraints\footnote{One can first consider the cases $I=J$ with $K$ odd, for which \eqref{constraint_Abis_Bianchi} implies $\A_{IKI}=0$ and thus $\A_{IIK}=0$. This yields the 4 constraints $\A_{001}=\A_{003}=\A_{113}=\A_{223}=0$. Let us now consider the cases $I\neq J$. Since all the constraints are symmetric in the exchange of $I$ and $J$, one can assume $I<J$. If $I+J$ is odd, and $K$ is thus even, one obtains in principle 6 relations, but 2 are redundant with the previous ones, so only 4 remain: $\A_{120}=\A_{012}$, $\A_{302}=\A_{023}$, $\A_{201}=-\A_{102}$, $\A_{212}=0$. Finally, if $I+J$ is even, and thus $K$ odd, we get 4 relations but 2 are redundant, so only 2 remain: $\A_{203}=-\A_{023}$ and $\A_{313}=0$.} between the $\A$ terms themselves, leaving 22 independent $\A$ terms.

\medskip
The Bianchi identities also provide some relations between the 1-limb sharing terms. 
Indeed, starting from the $\B$ term and effecting a ``rotation'' of the second vertex, one gets the relation
\beq
\B_{IJK}-(-1)^K\B_{IJK}+(-1)^J\Bter_{IJK}=0\,.
\eeq
Since $K$ must be odd to have a nonzero $\Bter_{IJK}$, this implies that all the nonzero $\Bter$ terms  can be expressed as functions of the $\B$, according to the expression
\beq
\label{Bter_B}
 \Bter_{IJK}=-2(-1)^J\B_{IJK}\quad \text{for} \;\; K \ {\rm odd} \, .
\eeq
Similarly, a rotation of the first vertex in the expression of $\Bbis_{IJK}$ yields
\beq
\Bbis_{IJK}+(-1)^J\Bter_{IJK}-\Bter_{IJK}=0\,.
\eeq
For $J$ even, this relation does not bring anything new, since we already know that $\Bbis_{IJK}=0$ in this case. For $J$ odd, we get $\Bbis_{IJK}=2\Bter_{IJK}$, which implies $\Bbis_{IJK}=0$ for $K$ even, which we knew already. Finally, in the case where $J$ and $K$ are odd, we obtain, after substitution of \eqref{Bter_B}, the new relation
\beq
\Bbis_{IJK}=4\B_{IJK}\quad {\rm for} \ J \ {\rm and}\ K \  {\rm odd}\,,
\eeq
which thus determines all the nonzero  $\Bbis$ in terms of the $\B$.

In summary, taking into account the index symmetries and the $70$ independent Bianchi identities, we have found that all  possible terms of the form (\ref{DFDF}) can be expressed as combinations of 22 $\A$ terms and 32 $\B$ terms.

\subsection{Dimensionally dependent identities}

We now consider scalar DDIs of the form (\ref{DDI_scalar}) that provide relations between the $\A$ and $\B$ families of terms.
To obtain such terms, the tensor $\Lambda$  in (\ref{DDI_scalar}) must contain two derivatives of $F$ and arbitrary  combinations of  $F_I$ with possibly internal  contractions  such that the whole tensor possesses 10 free indices. 
In other words,  the $\Lambda$ tensor can be seen as a (tensor) product of two vertices and  of several $F_I$. The systematic exploration of all 
 possible $\Lambda$ tensors and corresponding DDI relations is a bit involved and the details are explained in the appendix. 
 
 \medskip 
 Many of these identities, especially when the Bianchi identities are taken into account, turn out to be redundant. We have thus identified  32 independent DDIs, which can be obtained from a single type of expression for the $\Lambda$ tensor, which we give here explicitly.  In order to express the $\Lambda$ tensor with ten contravariant indices, we introduce the partially antisymmetric tensor 
 \beq
 \Delta_{a_1a_2a_3a_4a_5 b_1b_2b_3b_4b_5}=\delta_{a_1a_2a_3a_4a_5}^{c_1c_2c_3c_4c_5}\prod_{n=1}^5 g_{c_nb_n}, \label{PAST}
 \eeq
so that the scalar DDIs (\ref{DDI_scalar}) are now written in the form
\beq
 \Delta_{a_1a_2a_3a_4a_5 b_1b_2b_3b_4b_5} \Lambda^{a_1a_2a_3a_4a_5 b_1b_2b_3b_4 b_5}=0\,.
\eeq
Let us also introduce  the ``dressed'' vertex tensors
\beq
T^{abc}_{IJK}\equiv F_I^{ae}F_J^{bf}F_K^{cg}\nabla_e F_{fg}\,,
\eeq
which are useful for the systematic investigation of all possible DDIs, as discussed in the appendix. The only family of relevant DDI identities can then be obtained from 
\beq
\mathcal{D}_{IJKL}\equiv  \Delta_{a_1a_2a_3a_4a_5 b_1b_2b_3b_4b_5} F^{b_4a_4}F^{b_5a_5} T^{a_1 a_2 b_1}_{KLI} T^{b_2 b_3 a_3}_{J00}=0\,. \label{L3IIC2}
 \eeq
As said above, only 32 DDIs of this class can cover them all once the Bianchi identity is taken into account. They can be chosen to be :
 \beq
 \left\{
\begin{matrix}
\mathcal{D}_{0000} &  \mathcal{D}_{0100} & \mathcal{D}_{1000} & \mathcal{D}_{0002} & \mathcal{D}_{0101} & \mathcal{D}_{0110} & \mathcal{D}_{0200} & \mathcal{D}_{1001}  \\
\mathcal{D}_{1010} & \mathcal{D}_{1100} & \mathcal{D}_{2000} & \mathcal{D}_{2001} & \mathcal{D}_{2010} & \mathcal{D}_{2100} & \mathcal{D}_{3000} & \mathcal{D}_{1210} \\
\mathcal{D}_{1300} &  \mathcal{D}_{2002} & \mathcal{D}_{2011} & \mathcal{D}_{2020} & \mathcal{D}_{2101} & \mathcal{D}_{2110} & \mathcal{D}_{2200} & \mathcal{D}_{3001} \\
\mathcal{D}_{3010} & \mathcal{D}_{3100} & \mathcal{D}_{3110} & \mathcal{D}_{3200} & \mathcal{D}_{3030} & \mathcal{D}_{3201} & \mathcal{D}_{3210} & \mathcal{D}_{3300}\
\end{matrix}
\right\}
 \eeq 
To give an explicit example, the first DDI of the list above, $\mathcal{D}_{0000}$,   gives the relation 
 \bea
\f \A_{000}+4\A_{002}+2\A_{011}-4\A_{101}+2\A_{200}-\f \At_{000}-2\At_{002}-4\At_{020}+2\At_{110}
\\ \nonumber
\qquad
 +\f \B_{000}+4\B_{002}+2\B_{011}-4\B_{101}+8\B_{110}+2\B_{200} +2\Bbis_{011}-4\Bter_{011}+4\Bter_{101}=0\,.
\eea
After simplification using (\ref{constraint_Abis_Bianchi}), this reduces to
\bea
2\A_{011}-4 \A_{002}-2\A_{200}-\f \A_{000}+2\B_{011}
+4\B_{002}-4\B_{101}+2\B_{200}+\f \B_{000}
=0\,,
\eea
which provides a new relation between the $\A$ and $\B$ of weight 2.

\medskip
In summary, the 54 remaining Lagrangian terms quadratic in $\nabla F$, which include  22 $\A$ terms and 32 $\B$ terms can be further reduced to only 22 ``seed'' terms, since all the other terms  can be recovered from these ones. The choice of these 22 terms, together with the four Riemann terms obtained in the previous section, constitutes a choice of basis. There is some flexibility in the choice of the 22 representative terms, but it does not seem possible to eliminate all the $\A$ or all the $\B$. 

%%%%%%%%%%%%%%
\section{Boundary terms}
%%%%%%%%%%%%%%
\label{section_boundary}
We now study whether it is possible to relate some of the 26 terms of the algebraic basis obtained so far,  via  boundary terms, since the latter are irrelevant in the equations of motion. Boundary terms are of the form $\nabla_a B^a$ where $B$ is some vector. In our case, we are interested in boundary terms $\nabla_a B^a$ that lead, when expanded,   to terms quadratic in  $\nabla F$, discussed in the previous section,  and therefore  $B$ must contain a single derivative of $F$. 

\subsection{Dressed vertex vectors}

Vectors $B$  of this form are proportional to one of the following contracted dressed vertices:
\beq
\label{def_V}
 \V^a_{IJ}\equiv F_I^{ce}F_J^{ad}\nabla_e F_{cd}
 \eeq
 or
\beq
\label{def_U}
\U^a_{IJ}\equiv F_I^{dc}F_J^{ae}\nabla_e F_{cd}=\frac{1}{I+1}F_J^{ae}\nabla_e( F^d_{I+1, d})
\eeq
where the last  expression shows that $\U_{IJ}$  vanishes if $I$ is even.

Replacing $\nabla_e F_{cd}$ in (\ref{def_U}), or equivalently in (\ref{def_V}), by $\nabla_{[e} F_{cd]}=0$ yields the relation
\beq
\V_{IJ}-(-1)^I \V_{IJ}+\U_{IJ}=0\,,
\eeq
which implies 
\beq
\label{U_V}
\U_{IJ}=-2 \V_{IJ} \quad {\rm for}\  I\, {\rm odd}\,, \qquad \U_{IJ}=0\quad {\rm for}\ I\, {\rm even}\,,
\eeq
where we have omitted the spacetime index for all vector terms above  to simplify the notation.
 As a consequence, we need to consider only the vectors $ \V_{IJ}$  since all the non-zero $\U_{IJ}$ can be expressed in terms of them.

 \subsection{Divergence of the vertex vectors}
  Let us now examine, for the above vectors, in which case their  divergence leads to only ``permissible" terms. In other words, after the full expansion of the vector $B$, one must find only terms of the form  (\ref{DFDF}) and, for example,  no term of the form $\nabla_a\nabla_b F_{cd}$ which cannot be reduced to a Riemann term.

For $I$ odd, the divergence of
\beq
\V^a_{IJ} =-\frac12 \U^a_{IJ} =-\frac{1}{2(I+1)}F_J^{ae}\nabla_e( F^d_{I+1, d})
\eeq
leads to 
\beq
\nabla_a \V^a_{IJ} =-\frac{1}{2(I+1)}\left[(\nabla_aF_J^{ae})\nabla_e( F^d_{I+1, d}) +F_J^{ae}\nabla_a\nabla_e( F^d_{I+1, d})\right]\,
\eeq
where the  last term on the right-hand side, which is not permissible, vanishes if $I$ is odd. Therefore, each of the four terms $\V_{11}$, $\V_{13}$, $\V_{31}$ and $\V_{33}$ constitutes an appropriate boundary term.

\medskip 
For the other values of  $I$ and $J$, it is instructive to expand  the divergence of $\V_{IJ}$:
\bea
 \nabla_a\V^a_{IJ}= F_I^{ce}F_J^{ad}\nabla_a\nabla_e F_{cd}+F_I^{ce}(\nabla_aF_J^{ad})\nabla_e F_{cd}+F_J^{ad}(\nabla_aF_I^{ce})\nabla_e F_{cd}
\eea
The first term on the right hand side, which is not permissible, can however be rewritten as
\beq
F_{I}^{ce} F_{J}^{ad}\nabla_a \nabla_e F_{cd}=F_{I}^{ce} F_{J}^{ad}[\nabla_a, \nabla_e] F_{cd}+F_{I}^{ce} F_{J}^{ad}\nabla_e \nabla_a F_{cd}\,.
\eeq
The last term on the right hand side can be slightly rewritten, after exchanging the index labels $a$ and $e$  and the labels $c$ and $d$,  so that one gets
\beq
F_{I}^{ce} F_{J}^{ad}\nabla_a \nabla_e F_{cd}=F_{I}^{ce} F_{J}^{ad}[\nabla_a, \nabla_e] F_{cd}-(-1)^{I+J}F_{J}^{ce} F_{I}^{ad}\nabla_a \nabla_e F_{cd}
\eeq
Reexpressing the commutator of the covariant derivatives in terms of the Riemann tensor, one finally obtains the relation
\beq
F_{I}^{ce} F_{J}^{ad}\nabla_a \nabla_e F_{cd}+(-1)^{I+J}F_{J}^{ce} F_{I}^{ad}\nabla_a \nabla_e F_{cd}
=F_{I}^{ce} F_{J}^{ad}(R_{eac}^{\ \ \ f} F_{df}+R_{ead}^{\ \ \ f} F_{fc})
\eeq
This shows that even if a vertex vector with 
$I$ even
is not an acceptable boundary term, the  combination of  two such vertices
\beq
\W_{IJ}\equiv \V_{IJ}+(-1)^{I+J} \V_{JI}
\eeq
provides an adequate boundary term. We thus find 7 additional boundary terms: $\W_{00}$, $\W_{02}$, $\W_{22}$, $\W_{01}$, $\W_{03}$, $\W_{12}$ and $\W_{23}$.
In total, we have thus identified 11 boundary terms. However, as we are going to see below, some of these are redundant.

\subsection{Divergenceless vectors}
One can find relations between some of the 11 boundary terms obtained above, as we now show. 
One such relation comes from  the identity
\beq
\nabla_a\nabla_b\left(\Phi F^{ab}\right)=0 \,,
\eeq
where $\Phi$ is a function of the two invariants $\f$ and $f_4$, which implies that the vector 
\beq
\nabla_b\left(\Phi F^{ab}\right)=\Phi \nabla_b F^{ab}+2(\partial_2\Phi) F^{ab}F^{cd}\nabla_b F_{cd}+4(\partial_4\Phi) F^{ab} F_3^{cd}\nabla_b F_{cd}\,,
\eeq
is divergenceless. As a consequence, any boundary  term constructed from $\V_{00}^a=-\nabla_b F^{ab}$ can be considered  as redundant since it can be replaced by other terms, according to the relation
\beq
\nabla_a(\Phi \V_{00}^a)= \nabla_a\left[4(\partial_2\Phi)\V_{11}^a+8 (\partial_4\Phi) \V_{31}^a \right]\,,
\eeq
where we have used (\ref{U_V}) to replace the $\U$ terms on the right hand side by $\V$ terms.

Similarly, the identity
\beq
\nabla_a\nabla_b\left(\Phi F_3^{ab}\right)=0 
\eeq
yields the relation
\beq
\nabla_a\left(\Phi \V_{02}^a+\Phi\V_{11}^a+\Phi \V_{20}^a\right)=\nabla_a\left[4(\partial_2\Phi)\V_{31}^a+8 (\partial_4\Phi)\V_{33}^a \right)]\,,
\eeq
which implies that one term among the three terms $\V_{02}$,  $\V_{11}$ and $\V_{20}$ is redundant.

\subsection{DDIs for the boundary terms}
Finally, one can also find a few non trivial DDIs that provide additional relations between the boundary terms.
They are  given by
\beq
 \Delta_{a_1a_2a_3a_4a_5 b_1b_2b_3b_4b_5}\, \nabla^{b_2}\!\left( \Phi T^{a_1a_2b_1}_{00I}  F^{b_3b_4} F^{a_3a_4} F^{b_5a_5} \right) \propto \nabla^e \left( \Phi F_I^{ab} \nabla_a C_{be} \right)= 0 \,.
\eeq
where $C$ is the left-hand-side of the Cayley-Hamilton identity \eqref{CH}. They provide four relations between the $\V_{IJ}$ and the $\U_{IJ}$. After the replacement of the $\U_{IJ}$ in terms of the $\V_{IJ}$, one gets the following relations
\bea
\W_{03}-\W_{12}+\frac12 \f \W_{01}=0\,,
\\
\V_{13}+\V_{31}-\frac12\W_{22}+\frac12 \f \V_{11}-\frac18 \g \W_{00}=0\,,
\\
\W_{23}+\frac14 \g \W_{01}=0\,,
\\
\V_{33}+\frac14 \f\W_{22}+\frac14 \g (\V_{11}-\W_{02})=0\,.
\eea
In summary, the four above relations combined with the two relations discussed in the previous subsection means that among the 11 initial boundary terms, only  5 boundary terms are independent.

\section{Basis of elementary Lagrangians}
\label{section_basis}
Thanks to the results of the previous sections, we are now able to rewrite any action of the form  \eqref{action} in terms of a linear combination of 21 terms (not including $L_0$), with coefficients that are arbitrary functions of $\f$ and $f_4$. One possible choice of these 21 terms is the following: the Riemann-linear terms are
\begin{eqnarray}
\mathcal{R}_1\equiv R  \;,\;\;\;\;
\mathcal{R}_2\equiv F^{ab} F^{cd} R_{abcd} \;,\;\;\;\;\mathcal{R}_3\equiv F_{2}^{ab} R_{ab}\,,
\end{eqnarray}
while  the 18 terms quadratic in $\nabla\F$ chosen among the scalars $\A_{IJK}$ and $\B_{IJK}$ and  organised according to their weight (i.e. the sum $I+J+K$), comprise  one term
of weight 0,
\begin{eqnarray}
\mathcal{F}_{1}\equiv  \B_{000} = \nabla^bF_{ab}\nabla^cF_{c}{}^{\, a} \, , 
\end{eqnarray}
seven terms of weight 2,
\begin{eqnarray}
\begin{array}{ll}
\mathcal{F}_2\equiv  \A_{002}  = F_2^{ec} \nabla_e F_{ab} \nabla^b F_{c}^{\ a}\, , & \quad
\mathcal{F}_{3}\equiv \A_{200} = F_2^{ad} \nabla^c F_{ab} \nabla^b F_{cd}  \, ,\\
\mathcal{F}_{4} \equiv  \B_{020} = F_2^{eb} \nabla_e F_{ab} \nabla_c F^{ca} \, ,  & \quad 
\mathcal{F}_{5} \equiv \B_{200} = F_2^{ad} \nabla^b F_{ab} \nabla^c F_{cd} \, , \\
\mathcal{F}_6 \equiv  \A_{101}  = F^{ad} F^{ec}  \nabla_e F_{ab} \nabla^b F_{cd}  \, , & \quad
\mathcal{F}_{7} \equiv \B_{011} = F^{eb} F^{cf} \nabla_e F_{ab} \nabla_f F_{c}^{\ a}  \, , \\
\mathcal{F}_{8}\equiv \B_{110} = F^{ad}F^{eb}  \nabla_e F_{ab} \nabla^c F_{cd}  \, , &
\end{array}
\end{eqnarray}
one term of weight 3,
\begin{eqnarray}
\mathcal{F}_{9} \equiv \B_{210} = F_2^{ad}F^{eb} \nabla_e F_{ab} \nabla^c F_{cd}  \, , 
\end{eqnarray}
six terms of weight 4,
\begin{eqnarray}
\begin{array}{ll}
\mathcal{F}_{10}\equiv  \A_{301} = F_3^{ad} F^{ec}  \nabla_e F_{ab} \nabla^b F_{cd}   \, ,  &\quad
\mathcal{F}_{11}\equiv \B_{310}= F_3^{ad}F^{eb} \nabla_e F_{ab} \nabla^c F_{cd}  \, ,  \\
\mathcal{F}_{12}\equiv \B_{022} = F_2^{eb} F_2^{cf} \nabla_e F_{ab} \nabla_f F_{c}^{\ a} \, ,  &\quad
\mathcal{F}_{13} \equiv \B_{220}  = F_2^{ad}F_2^{eb}  \nabla_e F_{ab} \nabla^c F_{cd}  \, ,  \\
\mathcal{F}_{14}\equiv \B_{112} = F^{ad}F^{eb} F_2^{cf} \nabla_e F_{ab} \nabla_f F_{cd}   \, ,  & \quad
\mathcal{F}_{15} \equiv \B_{211}= F_2^{ad}F^{eb} F^{cf} \nabla_e F_{ab} \nabla_f F_{cd}  \, ,
\end{array}
\end{eqnarray}
two terms of weight 5,
\begin{eqnarray}
\begin{split}
\mathcal{F}_{16}\equiv \B_{320}  = F_3^{ad}F_2^{eb} \nabla_e F_{ab} \nabla^c F_{cd} \, ,\quad
\mathcal{F}_{17}\equiv \B_{212} = F_2^{ad}F^{eb} F_2^{cf} \nabla_e F_{ab} \nabla_f F_{cd} \, ,
\end{split}
\end{eqnarray}
and one last term of weight 6,
\begin{eqnarray}
\mathcal{F}_{18} \equiv \B_{312} = F_3^{ad}F^{eb} F_2^{cf} \nabla_e F_{ab} \nabla_f F_{cd} \,.
\end{eqnarray}

Note that the Riemann-linear term $\R_4$, defined in \eqref{basis_Riemann}, does not appear in the basis. Indeed, using the relations obtained in the previous section from the boundary terms, this term can be rewritten as a linear combination of the terms in the above basis. Explicitly, one finds
 \begin{eqnarray}
 \begin{split}
\Phi \mathcal{R}_4 &=\Phi \left(2 \mathcal{F}_{6} - \frac12 g_4 \R_1  \right) + f_2 \, g_4\, (\partial_4 \Phi) \mathcal{R}_2 + 2 \s_1 \left(\frac12{f_2}\, \mathcal{R}_3 - \mathcal{F}_{2}- \mathcal{F}_{3} + \mathcal{F}_{4} +  \mathcal{F}_{5} \right) \\
&+ 8 f_2 \, \s_2 \left( 2 \mathcal{F}_{11} - \mathcal{F}_{13}-\frac14{  g_4} \mathcal{F}_{1} \right) + 8 \s_3 \left(  \mathcal{F}_{14} -  \mathcal{F}_{15} \right) -2 (\s_4- \s_5 )  \mathcal{F}_{7} + 2 (\s_4-2 \s_5)  \mathcal{F}_{8}\,,
\end{split}
 \end{eqnarray}
 where 
 \begin{eqnarray}
 \begin{split}
\s_1 &= \Phi + 2 g_4 (\partial_4 \Phi) \,,\;\;\; \s_2 = 3 (\partial_4 \Phi) + 2 g_4 (\partial_{44} \Phi) \,,\;\;\; \s_3 =  \partial_2 \Phi + 4 f_2  (\partial_4 \Phi) +2 f_2 \, g_4  (\partial_{44} \Phi)  \\
 \s_4 &= - \Phi + 2 f_2 \left[ \partial_{2}\Phi +2 g_4 (\partial_{24}\Phi) \right] + 2 \left( g_4 - 2 f_2^2 \right)\partial_{4}\Phi  ,\;\; \s_5 =  - \Phi - 2 f_2^2 \left[ 3 (\partial_{4}\Phi)+ 2 g_4 (\partial_{44}\Phi) \right].
 \end{split}
 \end{eqnarray}

In flat spacetime the term $\R_4$  vanishes  (as well as  the other $\R_i$)   and the above relation implies that one of the $\mathcal{F}_i$, for example  $\mathcal{F}_{6}$,  becomes redundant.  Hence, any action of the form  \eqref{action} in Minkowski can be written as a linear combination of 17 terms, all quadratic in  derivatives of $F_{ab}$, with  arbitrary functions of $\f$ and $f_4$ as coefficients.

Finally,  let us stress  that it is straightforward to adapt our analysis to  classify all the Lagrangians of the form  \eqref{action}  
where $F_{ab}$ is now an arbitrary 2-form, i.e. not associated with a $U(1)$ gauge field. The only difference is that one cannot use anymore the relations obtained from Bianchi identities (\ref{Bianchi_F}). 
The Riemann-linear terms \eqref{R_terms_generic} are the same  and can be reduced to a combination of the four terms (\ref{basis_Riemann}). As for the terms of the form \eqref{DFDF}, we again start with 124 terms,  as discussed in section \ref{familiesAandB}, and  use 88 independent DDIs to reduce these terms to 36 elementary ones. We then find that among the 14 possible 
boundary terms,  only 6 of them are independent when we 
consider their DDIs and the divergenceless vectors. We thus end up with  30 independent elementary quadratic Lagrangians of the form   \eqref{DFDF}. In conclusion, adding the 4 terms linear in the Riemann tensor, we get  a basis of 34 independent Lagrangians for the most general action  \eqref{action} when $F_{ab}$ is an arbitrary 2-form.

\section{Conclusions}
In this article, we have classified (and reduced)  higher-order Einstein-Maxwell theories described by Lagrangians that are linear in the Riemann tensor and quadratic in derivatives of the field strength tensor. As we have seen, classifying such Lagrangians turns out to be much more complex than for scalar-tensor theories. The main reason is that we need to take into account an important number of dimensionally dependent identities, which renders the analysis of the equivalence between the Lagrangians, when boundary terms are also included, much more subtle but also very interesting. 

Our results show that the most general higher-order Einstein-Maxwell  Lagrangians considered here can be written as a linear combination of 21 terms whose  coefficients are arbitrary functions of the two scalar invariants constructed from the field-strength tensor. We also provide an explicit basis. These results could also be useful to simplify, or more simply to compare, effective actions that contain terms of the form (\ref{action}), such as those appearing in the derivative expansion of the one-loop effective action of QED \cite{Gusynin:1998bt,Navarro-Salas:2020oew,Karbstein:2021obd}. Furthermore, generalising our approach to gravitational actions could be used in other contexts such as Asymptotic Safety (see e.g. \cite{Knorr:2019atm}, where non-perturbative effects are parametrised by invariants involving form-factors). As mentioned in the previous section, our analysis is readily extendible to the case where the Lagrangians are seen as functionals of a 2-form $F_{ab}$, thus relaxing the $U(1)$ gauge symmetry.

At this stage, we have not investigated the physical properties of the theories (\ref{action}), in particular the nature and the number of their degrees of freedom. As the Lagrangians involve 
second derivatives of the gauge field $A^a$, which   in general cannot be eliminated by integrations by part, they could contain unwanted Ostrogradski ghosts. 
To avoid these, one must look for a degenerate family of Lagrangians, satisfying  degeneracy conditions analogous to those that have been identified in DHOST theories, to ensure that the theory does not propagate Ostrogradski ghosts even though the equations of motion are higher order. 

Indeed, a simple preliminary analysis shows that theories of the form (\ref{action})  could propagate up to three Ostrogradski ghosts, in addition to the usual two degrees of freedom of the photon and the two degrees of freedom of the graviton. 
The problem of finding necessary and sufficient degeneracy conditions to get rid of these ghosts is much more involved than in higher-derivative scalar-tensor theories for the reason that a gauge field $A^a$ has a more complex structure than a scalar field.   A first exploration of these aspects will be considered for the flat case in a future work.

\acknowledgments
 This work was supported by the
French National Research Agency (ANR) via Grant No. ANR-22-CE31-0015-01 associated with the project StronG. A.C. was supported by the Research Authority of the Open University of Israel (grant number 512343). We acknowledge the use of the xAct package for Mathematica  \cite{martin2002xact} and Mathcha to draw the diagrams.

\appendix

\section{Dimensionally dependent identities}\label{app_DDI}

In this appendix we classify dimensionally dependent identities (DDIs) built from the quadratic combination $\nabla\F\nabla\F$ and arbitrary products of field strength tensors in four dimensions. Our purpose is to identify all the possible 10-index tensors which can appear in \eqref{DDI_scalar}. Given that there can be at most nine field strengths together with $\nabla\F\nabla\F$ in the corresponding scalar invariants, a rough estimate to enumerate possible DDIs is to choose 10 among 24 indices, resulting in around two million possibilities. Thus, it is desirable to obtain some systematic control over such problems.

\subsection{Diagrammatic classification}

In order to do so, we use a graphical representation of the various tensorial quantities involved in the scalar invariants of our classification. The coloured field strength is represented by
\begin{eqnarray}
\begin{tikzpicture}[x=0.75pt,y=0.75pt,yscale=-1,xscale=1]
\draw    (150.36,70.55) -- (127.12,61.53) ;
\draw    (127.14,61.52) -- (150.38,52.49) ; 
\draw (71.5,51.61) node [anchor=north west][inner sep=0.75pt]    {$F_{I}^{ab} \equiv $};
\draw (153.77,45.85) node [anchor=north west][inner sep=0.75pt]  [font=\tiny]  {$a$};
\draw (154.1,65.51) node [anchor=north west][inner sep=0.75pt]  [font=\tiny]  {$b$};
\draw (114.5,57.7) node [anchor=north west][inner sep=0.75pt]  [font=\tiny]  {$I$};
\end{tikzpicture}
\end{eqnarray}
while for clarity, we recall the definition of the coloured vertex tensor, given by
\begin{eqnarray}
\begin{tikzpicture}[x=0.75pt,y=0.75pt,yscale=-1,xscale=1]

%Straight Lines [id:da32884065145737607] 
\draw    (308.51,149.84) -- (285.27,140.81) ;
%Shape: Boxed Line [id:dp03055918797145618] 
\draw    (285.28,140.8) -- (308.52,131.77) ;
%Shape: Wave [id:dp507624685042084] 
\draw   (262.28,140.99) .. controls (262.52,141.37) and (262.74,141.73) .. (263,141.73) .. controls (263.26,141.73) and (263.49,141.37) .. (263.72,140.99) .. controls (263.96,140.61) and (264.18,140.25) .. (264.44,140.25) .. controls (264.7,140.25) and (264.93,140.61) .. (265.16,140.99) .. controls (265.4,141.37) and (265.62,141.73) .. (265.88,141.73) .. controls (266.14,141.73) and (266.37,141.37) .. (266.6,140.99) .. controls (266.84,140.61) and (267.06,140.25) .. (267.32,140.25) .. controls (267.58,140.25) and (267.81,140.61) .. (268.04,140.99) .. controls (268.28,141.37) and (268.5,141.73) .. (268.76,141.73) .. controls (269.02,141.73) and (269.25,141.37) .. (269.48,140.99) .. controls (269.72,140.61) and (269.94,140.25) .. (270.2,140.25) .. controls (270.46,140.25) and (270.69,140.61) .. (270.92,140.99) .. controls (271.16,141.37) and (271.38,141.73) .. (271.64,141.73) .. controls (271.91,141.73) and (272.13,141.37) .. (272.36,140.99) .. controls (272.6,140.61) and (272.82,140.25) .. (273.09,140.25) .. controls (273.35,140.25) and (273.57,140.61) .. (273.81,140.99) .. controls (274.04,141.37) and (274.27,141.73) .. (274.53,141.73) .. controls (274.79,141.73) and (275.01,141.37) .. (275.25,140.99) .. controls (275.48,140.61) and (275.71,140.25) .. (275.97,140.25) .. controls (276.23,140.25) and (276.45,140.61) .. (276.69,140.99) .. controls (276.92,141.37) and (277.15,141.73) .. (277.41,141.73) .. controls (277.67,141.73) and (277.89,141.37) .. (278.13,140.99) .. controls (278.36,140.61) and (278.59,140.25) .. (278.85,140.25) .. controls (279.11,140.25) and (279.33,140.61) .. (279.57,140.99) .. controls (279.8,141.37) and (280.03,141.73) .. (280.29,141.73) .. controls (280.55,141.73) and (280.77,141.37) .. (281.01,140.99) .. controls (281.24,140.61) and (281.47,140.25) .. (281.73,140.25) .. controls (281.99,140.25) and (282.21,140.61) .. (282.45,140.99) .. controls (282.68,141.37) and (282.91,141.73) .. (283.17,141.73) .. controls (283.43,141.73) and (283.65,141.37) .. (283.89,140.99) .. controls (284.12,140.61) and (284.35,140.25) .. (284.61,140.25) .. controls (284.87,140.25) and (285.09,140.61) .. (285.33,140.99) .. controls (285.33,140.99) and (285.33,140.99) .. (285.33,140.99) ;

% Text Node
\draw (62.5,128.9) node [anchor=north west][inner sep=0.75pt]    {$T_{IJK}^{abc} \equiv F_{I}^{ae} F_{J}^{bf} F_{K}^{cg} \nabla _{e} F_{fg} =$};
% Text Node
\draw (311.91,125.13) node [anchor=north west][inner sep=0.75pt]  [font=\tiny]  {$b,J$};
% Text Node
\draw (312.24,144.8) node [anchor=north west][inner sep=0.75pt]  [font=\tiny]  {$c,K$};
% Text Node
\draw (242.64,137.27) node [anchor=north west][inner sep=0.75pt]  [font=\tiny]  {$a,I$};
\end{tikzpicture}\label{DefTtensor}
\end{eqnarray}
Although we could introduce an orientation on these graphs in order to encode the positions of antisymmetrised indices with respect to that of the covariant derivative, we will refrain to do so, our purpose here being a classification rather than actual calculations of DDIs.

The previous tensorial quantities are sufficient to define all the DDIs, however it is quite convenient to construct other coloured tensors from these building blocks. In particular, we construct and represent the traces of $T$ by
\begin{eqnarray}
\begin{tikzpicture}[x=0.75pt,y=0.75pt,yscale=-1,xscale=1]

%Shape: Circle [id:dp7987781781721456] 
\draw  [fill={rgb, 255:red, 0; green, 0; blue, 0 }  ,fill opacity=0.99 ] (196.14,123.93) .. controls (196.14,122.57) and (197.24,121.47) .. (198.6,121.47) .. controls (199.95,121.47) and (201.05,122.57) .. (201.05,123.93) .. controls (201.05,125.28) and (199.95,126.38) .. (198.6,126.38) .. controls (197.24,126.38) and (196.14,125.28) .. (196.14,123.93) -- cycle ;
%Straight Lines [id:da007911802524084521] 
\draw    (198.6,123.93) -- (226.17,124.01) ;

% Text Node
\draw (82,112.53) node [anchor=north west][inner sep=0.75pt]    {$V_{IJ}^{\ a} \equiv g_{bc} T_{I0J}^{bca} =$};
% Text Node
\draw (229.17,119.4) node [anchor=north west][inner sep=0.75pt]  [font=\tiny]  {$a,J$};
% Text Node
\draw (195,112.35) node [anchor=north west][inner sep=0.75pt]  [font=\tiny]  {$I$};

\end{tikzpicture}
\end{eqnarray}
\begin{eqnarray}
\begin{tikzpicture}[x=0.75pt,y=0.75pt,yscale=-1,xscale=1]

%Shape: Wave [id:dp42343680627246183] 
\draw   (180.01,103.78) .. controls (180.26,104.2) and (180.5,104.59) .. (180.78,104.59) .. controls (181.06,104.59) and (181.3,104.2) .. (181.55,103.78) .. controls (181.8,103.36) and (182.04,102.97) .. (182.32,102.97) .. controls (182.6,102.97) and (182.84,103.36) .. (183.09,103.78) .. controls (183.34,104.2) and (183.58,104.59) .. (183.86,104.59) .. controls (184.13,104.59) and (184.37,104.2) .. (184.62,103.78) .. controls (184.87,103.36) and (185.11,102.97) .. (185.39,102.97) .. controls (185.67,102.97) and (185.91,103.36) .. (186.16,103.78) .. controls (186.41,104.2) and (186.65,104.59) .. (186.93,104.59) .. controls (187.21,104.59) and (187.45,104.2) .. (187.7,103.78) .. controls (187.95,103.36) and (188.19,102.97) .. (188.47,102.97) .. controls (188.75,102.97) and (188.99,103.36) .. (189.24,103.78) .. controls (189.49,104.2) and (189.73,104.59) .. (190.01,104.59) .. controls (190.28,104.59) and (190.52,104.2) .. (190.77,103.78) .. controls (191.02,103.36) and (191.26,102.97) .. (191.54,102.97) .. controls (191.82,102.97) and (192.06,103.36) .. (192.31,103.78) .. controls (192.56,104.2) and (192.8,104.59) .. (193.08,104.59) .. controls (193.36,104.59) and (193.6,104.2) .. (193.85,103.78) .. controls (194.1,103.36) and (194.34,102.97) .. (194.62,102.97) .. controls (194.9,102.97) and (195.14,103.36) .. (195.39,103.78) .. controls (195.64,104.2) and (195.88,104.59) .. (196.16,104.59) .. controls (196.43,104.59) and (196.67,104.2) .. (196.92,103.78) .. controls (197.17,103.36) and (197.41,102.97) .. (197.69,102.97) .. controls (197.97,102.97) and (198.21,103.36) .. (198.46,103.78) .. controls (198.71,104.2) and (198.95,104.59) .. (199.23,104.59) .. controls (199.51,104.59) and (199.75,104.2) .. (200,103.78) .. controls (200.25,103.36) and (200.49,102.97) .. (200.77,102.97) .. controls (201.05,102.97) and (201.29,103.36) .. (201.54,103.78) .. controls (201.79,104.2) and (202.03,104.59) .. (202.31,104.59) .. controls (202.58,104.59) and (202.82,104.2) .. (203.07,103.78) .. controls (203.33,103.36) and (203.56,102.97) .. (203.84,102.97) .. controls (204.12,102.97) and (204.36,103.36) .. (204.61,103.78) .. controls (204.61,103.78) and (204.61,103.78) .. (204.61,103.78) ;
%Shape: Circle [id:dp40796490181562506] 
\draw  [fill={rgb, 255:red, 0; green, 0; blue, 0 }  ,fill opacity=0.99 ] (176.14,103.93) .. controls (176.14,102.57) and (177.24,101.47) .. (178.6,101.47) .. controls (179.95,101.47) and (181.05,102.57) .. (181.05,103.93) .. controls (181.05,105.28) and (179.95,106.38) .. (178.6,106.38) .. controls (177.24,106.38) and (176.14,105.28) .. (176.14,103.93) -- cycle ;

% Text Node
\draw (62,92.53) node [anchor=north west][inner sep=0.75pt]    {$\tilde{V}_{IJ}^{\ a} \equiv g_{bc} T_{JI0}^{abc} =$};
% Text Node
\draw (209.17,99.4) node [anchor=north west][inner sep=0.75pt]  [font=\tiny]  {$a,J$};
% Text Node
\draw (175,92.35) node [anchor=north west][inner sep=0.75pt]  [font=\tiny]  {$I$};

\end{tikzpicture}
\end{eqnarray}
and we will represent by a spring either case
\begin{eqnarray}
\begin{tikzpicture}[x=0.75pt,y=0.75pt,yscale=-1,xscale=1]

%Shape: Spring [id:dp75062577490408] 
\draw   (94.6,33.45) .. controls (95.13,33.07) and (95.94,32.8) .. (97.1,32.8) .. controls (101.04,32.8) and (101.04,35.94) .. (99.07,35.94) .. controls (97.1,35.94) and (97.1,32.8) .. (101.04,32.8) .. controls (104.99,32.8) and (104.99,35.94) .. (103.02,35.94) .. controls (101.04,35.94) and (101.04,32.8) .. (104.99,32.8) .. controls (108.93,32.8) and (108.93,35.94) .. (106.96,35.94) .. controls (104.99,35.94) and (104.99,32.8) .. (108.93,32.8) .. controls (112.88,32.8) and (112.88,35.94) .. (110.91,35.94) .. controls (108.93,35.94) and (108.93,32.8) .. (112.88,32.8) .. controls (116.82,32.8) and (116.82,35.94) .. (114.85,35.94) .. controls (112.88,35.94) and (112.88,32.8) .. (116.82,32.8) .. controls (117.82,32.8) and (118.57,33) .. (119.1,33.3) ;
%Shape: Circle [id:dp5458971347234022] 
\draw  [fill={rgb, 255:red, 0; green, 0; blue, 0 }  ,fill opacity=0.99 ] (220.94,34.33) .. controls (220.94,32.97) and (222.04,31.87) .. (223.4,31.87) .. controls (224.75,31.87) and (225.85,32.97) .. (225.85,34.33) .. controls (225.85,35.68) and (224.75,36.78) .. (223.4,36.78) .. controls (222.04,36.78) and (220.94,35.68) .. (220.94,34.33) -- cycle ;
%Straight Lines [id:da21003598142420377] 
\draw    (223.4,34.33) -- (250.97,34.41) ;
%Shape: Wave [id:dp9330309186791339] 
\draw   (155.21,34.38) .. controls (155.46,34.8) and (155.7,35.19) .. (155.98,35.19) .. controls (156.26,35.19) and (156.5,34.8) .. (156.75,34.38) .. controls (157,33.96) and (157.24,33.57) .. (157.52,33.57) .. controls (157.8,33.57) and (158.04,33.96) .. (158.29,34.38) .. controls (158.54,34.8) and (158.78,35.19) .. (159.06,35.19) .. controls (159.33,35.19) and (159.57,34.8) .. (159.82,34.38) .. controls (160.07,33.96) and (160.31,33.57) .. (160.59,33.57) .. controls (160.87,33.57) and (161.11,33.96) .. (161.36,34.38) .. controls (161.61,34.8) and (161.85,35.19) .. (162.13,35.19) .. controls (162.41,35.19) and (162.65,34.8) .. (162.9,34.38) .. controls (163.15,33.96) and (163.39,33.57) .. (163.67,33.57) .. controls (163.95,33.57) and (164.19,33.96) .. (164.44,34.38) .. controls (164.69,34.8) and (164.93,35.19) .. (165.21,35.19) .. controls (165.48,35.19) and (165.72,34.8) .. (165.97,34.38) .. controls (166.22,33.96) and (166.46,33.57) .. (166.74,33.57) .. controls (167.02,33.57) and (167.26,33.96) .. (167.51,34.38) .. controls (167.76,34.8) and (168,35.19) .. (168.28,35.19) .. controls (168.56,35.19) and (168.8,34.8) .. (169.05,34.38) .. controls (169.3,33.96) and (169.54,33.57) .. (169.82,33.57) .. controls (170.1,33.57) and (170.34,33.96) .. (170.59,34.38) .. controls (170.84,34.8) and (171.08,35.19) .. (171.36,35.19) .. controls (171.63,35.19) and (171.87,34.8) .. (172.12,34.38) .. controls (172.37,33.96) and (172.61,33.57) .. (172.89,33.57) .. controls (173.17,33.57) and (173.41,33.96) .. (173.66,34.38) .. controls (173.91,34.8) and (174.15,35.19) .. (174.43,35.19) .. controls (174.71,35.19) and (174.95,34.8) .. (175.2,34.38) .. controls (175.45,33.96) and (175.69,33.57) .. (175.97,33.57) .. controls (176.25,33.57) and (176.49,33.96) .. (176.74,34.38) .. controls (176.99,34.8) and (177.23,35.19) .. (177.51,35.19) .. controls (177.78,35.19) and (178.02,34.8) .. (178.27,34.38) .. controls (178.53,33.96) and (178.76,33.57) .. (179.04,33.57) .. controls (179.32,33.57) and (179.56,33.96) .. (179.81,34.38) .. controls (179.81,34.38) and (179.81,34.38) .. (179.81,34.38) ;
%Shape: Circle [id:dp2193408931905274] 
\draw  [fill={rgb, 255:red, 0; green, 0; blue, 0 }  ,fill opacity=0.99 ] (151.34,34.53) .. controls (151.34,33.17) and (152.44,32.07) .. (153.8,32.07) .. controls (155.15,32.07) and (156.25,33.17) .. (156.25,34.53) .. controls (156.25,35.88) and (155.15,36.98) .. (153.8,36.98) .. controls (152.44,36.98) and (151.34,35.88) .. (151.34,34.53) -- cycle ;
%Shape: Circle [id:dp655589311578122] 
\draw  [fill={rgb, 255:red, 0; green, 0; blue, 0 }  ,fill opacity=0.99 ] (90.54,33.73) .. controls (90.54,32.37) and (91.64,31.27) .. (93,31.27) .. controls (94.35,31.27) and (95.45,32.37) .. (95.45,33.73) .. controls (95.45,35.08) and (94.35,36.18) .. (93,36.18) .. controls (91.64,36.18) and (90.54,35.08) .. (90.54,33.73) -- cycle ;

% Text Node
\draw (123.3,28.8) node [anchor=north west][inner sep=0.75pt]    {$=$};
% Text Node
\draw (191.3,27) node [anchor=north west][inner sep=0.75pt]   [align=left] {or};
% Text Node
\draw (54.1,26.4) node [anchor=north west][inner sep=0.75pt]    {$S\equiv $};

\end{tikzpicture}
\end{eqnarray}
so that $S=\U$ or $\V$. Let us remark that the elements of the two families introduced in the section \ref{familiesAandB} can easily be reconstructed from these definitions,
\bea
g_{ae}\, g_{bf}\, g_{cd}T^{abc}_{000}T^{def}_{JKI}= \A_{IJK} \,, \;\;\; g_{ad}\, g_{bf}\, g_{ce}T^{abc}_{000}T^{def}_{JKI}=\Abis_{IJK} \, ,
\eea
and
\bea
 g_{ab}\V^a_{IJ}\V^b_{K0}=-(-1)^{I+J}\B_{JIK}\,, \;\;\;g_{ab}\U^a_{IJ}\U^b_{K0}=(-1)^J\Bbis_{JIK}\,, \;\;\;g_{ab}\U^a_{IJ}\V^b_{K0}=-\Bter_{JKI}\, .
\eea

This enables to list all the disconnected diagrams representing coloured $\nabla\F\nabla\F$, built from tensorial products of the previous quantities, which are schematically of the form
\begin{eqnarray}
\left\{S^2, \V\U, ST, T^2\right\}\, .
\end{eqnarray}
Furthermore, we show in the penultimate section of this appendix that DDIs built from 2-index tensors, such as $\V\U$ and $S^2$, are redundant for both disconnected and connected diagrams. Therefore, we define only the connected diagrams representing coloured $\nabla\F\nabla\F$ with more than two indices. There are three families of these given by, 
\begin{eqnarray}
\begin{tikzpicture}[x=0.75pt,y=0.75pt,yscale=-1,xscale=1]

%Straight Lines [id:da5596875235981493] 
\draw    (238.79,49.5) -- (262.03,58.53) ;
%Shape: Wave [id:dp7933780247092714] 
\draw   (238.63,67.63) .. controls (238.92,67.93) and (239.2,68.21) .. (239.45,68.11) .. controls (239.7,68) and (239.86,67.54) .. (240.02,67.05) .. controls (240.18,66.56) and (240.34,66.1) .. (240.59,65.99) .. controls (240.84,65.88) and (241.12,66.17) .. (241.41,66.46) .. controls (241.71,66.76) and (241.99,67.04) .. (242.24,66.94) .. controls (242.49,66.83) and (242.64,66.37) .. (242.81,65.88) .. controls (242.97,65.39) and (243.12,64.93) .. (243.37,64.82) .. controls (243.63,64.72) and (243.91,65) .. (244.2,65.3) .. controls (244.49,65.59) and (244.77,65.88) .. (245.02,65.77) .. controls (245.27,65.66) and (245.43,65.2) .. (245.59,64.71) .. controls (245.75,64.22) and (245.91,63.76) .. (246.16,63.65) .. controls (246.41,63.55) and (246.69,63.83) .. (246.98,64.13) .. controls (247.28,64.42) and (247.56,64.71) .. (247.81,64.6) .. controls (248.06,64.5) and (248.21,64.03) .. (248.38,63.54) .. controls (248.54,63.06) and (248.69,62.59) .. (248.94,62.48) .. controls (249.2,62.38) and (249.48,62.66) .. (249.77,62.96) .. controls (250.06,63.26) and (250.34,63.54) .. (250.59,63.43) .. controls (250.84,63.33) and (251,62.86) .. (251.16,62.37) .. controls (251.32,61.89) and (251.48,61.42) .. (251.73,61.32) .. controls (251.98,61.21) and (252.26,61.49) .. (252.55,61.79) .. controls (252.85,62.09) and (253.13,62.37) .. (253.38,62.26) .. controls (253.63,62.16) and (253.78,61.69) .. (253.95,61.21) .. controls (254.11,60.72) and (254.26,60.25) .. (254.51,60.15) .. controls (254.77,60.04) and (255.05,60.32) .. (255.34,60.62) .. controls (255.63,60.92) and (255.91,61.2) .. (256.16,61.1) .. controls (256.41,60.99) and (256.57,60.53) .. (256.73,60.04) .. controls (256.89,59.55) and (257.05,59.09) .. (257.3,58.98) .. controls (257.55,58.87) and (257.83,59.16) .. (258.12,59.45) .. controls (258.42,59.75) and (258.7,60.03) .. (258.95,59.93) .. controls (259.2,59.82) and (259.35,59.36) .. (259.52,58.87) .. controls (259.68,58.38) and (259.83,57.92) .. (260.08,57.81) .. controls (260.34,57.71) and (260.62,57.99) .. (260.91,58.29) .. controls (261.2,58.58) and (261.48,58.87) .. (261.73,58.76) .. controls (261.87,58.7) and (261.97,58.54) .. (262.07,58.33) ;
%Straight Lines [id:da899781198568251] 
\draw    (262.03,58.53) -- (284.88,58.4) ;
%Straight Lines [id:da8051914987187546] 
\draw    (308.11,67.44) -- (284.87,58.41) ;
%Shape: Wave [id:dp6494033062131142] 
\draw   (308.26,49.31) .. controls (307.97,49.01) and (307.69,48.73) .. (307.44,48.83) .. controls (307.19,48.94) and (307.03,49.4) .. (306.87,49.89) .. controls (306.71,50.38) and (306.56,50.84) .. (306.3,50.95) .. controls (306.05,51.06) and (305.77,50.77) .. (305.48,50.48) .. controls (305.19,50.18) and (304.91,49.9) .. (304.66,50) .. controls (304.4,50.11) and (304.25,50.57) .. (304.09,51.06) .. controls (303.93,51.55) and (303.77,52.01) .. (303.52,52.12) .. controls (303.27,52.22) and (302.99,51.94) .. (302.69,51.64) .. controls (302.4,51.35) and (302.12,51.06) .. (301.87,51.17) .. controls (301.62,51.28) and (301.46,51.74) .. (301.3,52.23) .. controls (301.14,52.72) and (300.99,53.18) .. (300.73,53.29) .. controls (300.48,53.39) and (300.2,53.11) .. (299.91,52.81) .. controls (299.62,52.52) and (299.34,52.23) .. (299.09,52.34) .. controls (298.83,52.44) and (298.68,52.91) .. (298.52,53.4) .. controls (298.36,53.88) and (298.2,54.35) .. (297.95,54.46) .. controls (297.7,54.56) and (297.42,54.28) .. (297.12,53.98) .. controls (296.83,53.68) and (296.55,53.4) .. (296.3,53.51) .. controls (296.05,53.61) and (295.89,54.08) .. (295.73,54.57) .. controls (295.57,55.05) and (295.42,55.52) .. (295.16,55.62) .. controls (294.91,55.73) and (294.63,55.45) .. (294.34,55.15) .. controls (294.05,54.85) and (293.77,54.57) .. (293.52,54.68) .. controls (293.26,54.78) and (293.11,55.25) .. (292.95,55.73) .. controls (292.79,56.22) and (292.63,56.69) .. (292.38,56.79) .. controls (292.13,56.9) and (291.85,56.62) .. (291.55,56.32) .. controls (291.26,56.02) and (290.98,55.74) .. (290.73,55.84) .. controls (290.48,55.95) and (290.32,56.41) .. (290.16,56.9) .. controls (290,57.39) and (289.85,57.85) .. (289.59,57.96) .. controls (289.34,58.07) and (289.06,57.78) .. (288.77,57.49) .. controls (288.48,57.19) and (288.2,56.91) .. (287.95,57.01) .. controls (287.69,57.12) and (287.54,57.58) .. (287.38,58.07) .. controls (287.22,58.56) and (287.06,59.02) .. (286.81,59.13) .. controls (286.56,59.23) and (286.28,58.95) .. (285.98,58.65) .. controls (285.69,58.36) and (285.41,58.07) .. (285.16,58.18) .. controls (285.02,58.24) and (284.92,58.4) .. (284.82,58.61) ;

% Text Node
\draw (42,48.53) node [anchor=north west][inner sep=0.75pt]    {$X_{IJKLM}^{\ abcd} \equiv g_{ef} T_{JKI}^{abe} T_{LM0}^{cdf} =$};
% Text Node
\draw (219.17,62.73) node [anchor=north west][inner sep=0.75pt]  [font=\tiny]  {$a,J$};
% Text Node
\draw (218.84,42.73) node [anchor=north west][inner sep=0.75pt]  [font=\tiny]  {$b,K$};
% Text Node
\draw (311.51,42.73) node [anchor=north west][inner sep=0.75pt]  [font=\tiny]  {$c,L$};
% Text Node
\draw (311.84,62.4) node [anchor=north west][inner sep=0.75pt]  [font=\tiny]  {$d,M$};
% Text Node
\draw (269.67,49.16) node [anchor=north west][inner sep=0.75pt]  [font=\tiny]  {$I$};
\end{tikzpicture}
\end{eqnarray}
\begin{eqnarray}
\begin{tikzpicture}[x=0.75pt,y=0.75pt,yscale=-1,xscale=1]

%Straight Lines [id:da11917947429258358] 
\draw    (258.79,69.5) -- (282.03,78.53) ;
%Straight Lines [id:da8428112841027199] 
\draw    (328.11,87.44) -- (304.87,78.41) ;
%Shape: Boxed Line [id:dp6444205382842436] 
\draw    (282.03,78.53) -- (258.79,87.56) ;
%Shape: Boxed Line [id:dp3163923397215066] 
\draw    (304.88,78.4) -- (328.12,69.37) ;
%Shape: Wave [id:dp2602911594845211] 
\draw   (281.95,78.37) .. controls (282.18,78.75) and (282.41,79.11) .. (282.67,79.11) .. controls (282.93,79.11) and (283.15,78.75) .. (283.39,78.37) .. controls (283.62,77.99) and (283.85,77.63) .. (284.11,77.63) .. controls (284.37,77.63) and (284.59,77.99) .. (284.83,78.37) .. controls (285.06,78.75) and (285.29,79.11) .. (285.55,79.11) .. controls (285.81,79.11) and (286.03,78.75) .. (286.27,78.37) .. controls (286.5,77.99) and (286.73,77.63) .. (286.99,77.63) .. controls (287.25,77.63) and (287.48,77.99) .. (287.71,78.37) .. controls (287.95,78.75) and (288.17,79.11) .. (288.43,79.11) .. controls (288.69,79.11) and (288.92,78.75) .. (289.15,78.37) .. controls (289.39,77.99) and (289.61,77.63) .. (289.87,77.63) .. controls (290.13,77.63) and (290.36,77.99) .. (290.59,78.37) .. controls (290.83,78.75) and (291.05,79.11) .. (291.31,79.11) .. controls (291.57,79.11) and (291.8,78.75) .. (292.03,78.37) .. controls (292.27,77.99) and (292.49,77.63) .. (292.75,77.63) .. controls (293.01,77.63) and (293.24,77.99) .. (293.47,78.37) .. controls (293.71,78.75) and (293.93,79.11) .. (294.19,79.11) .. controls (294.45,79.11) and (294.68,78.75) .. (294.91,78.37) .. controls (295.15,77.99) and (295.37,77.63) .. (295.63,77.63) .. controls (295.89,77.63) and (296.12,77.99) .. (296.35,78.37) .. controls (296.59,78.75) and (296.81,79.11) .. (297.07,79.11) .. controls (297.33,79.11) and (297.56,78.75) .. (297.79,78.37) .. controls (298.03,77.99) and (298.25,77.63) .. (298.51,77.63) .. controls (298.77,77.63) and (299,77.99) .. (299.23,78.37) .. controls (299.47,78.75) and (299.69,79.11) .. (299.95,79.11) .. controls (300.21,79.11) and (300.44,78.75) .. (300.67,78.37) .. controls (300.91,77.99) and (301.13,77.63) .. (301.39,77.63) .. controls (301.66,77.63) and (301.88,77.99) .. (302.12,78.37) .. controls (302.35,78.75) and (302.57,79.11) .. (302.84,79.11) .. controls (303.1,79.11) and (303.32,78.75) .. (303.56,78.37) .. controls (303.79,77.99) and (304.02,77.63) .. (304.28,77.63) .. controls (304.54,77.63) and (304.76,77.99) .. (305,78.37) .. controls (305,78.37) and (305,78.37) .. (305,78.37) ;

% Text Node
\draw (62,68.53) node [anchor=north west][inner sep=0.75pt]    {$\tilde{X}_{IJKLM}^{\ abcd} \equiv g_{ef} T_{IJK}^{eab} T_{0LM}^{fcd} =$};
% Text Node
\draw (239.17,82.73) node [anchor=north west][inner sep=0.75pt]  [font=\tiny]  {$a,J$};
% Text Node
\draw (238.84,62.73) node [anchor=north west][inner sep=0.75pt]  [font=\tiny]  {$b,K$};
% Text Node
\draw (331.51,62.73) node [anchor=north west][inner sep=0.75pt]  [font=\tiny]  {$c,L$};
% Text Node
\draw (331.84,82.4) node [anchor=north west][inner sep=0.75pt]  [font=\tiny]  {$d,M$};
% Text Node
\draw (289.67,69.16) node [anchor=north west][inner sep=0.75pt]  [font=\tiny]  {$I$};

\end{tikzpicture}
\end{eqnarray}
\begin{eqnarray}
\begin{tikzpicture}[x=0.75pt,y=0.75pt,yscale=-1,xscale=1]

%Straight Lines [id:da14985540785385365] 
\draw    (278.79,50.17) -- (302.03,59.19) ;
%Straight Lines [id:da48752753757198464] 
\draw    (348.11,68.11) -- (324.87,59.08) ;
%Shape: Boxed Line [id:dp05068239617801462] 
\draw    (302.03,59.19) -- (278.79,68.22) ;
%Shape: Wave [id:dp8994197572396023] 
\draw   (301.95,59.37) .. controls (302.18,59.93) and (302.41,60.45) .. (302.67,60.45) .. controls (302.93,60.45) and (303.15,59.93) .. (303.39,59.37) .. controls (303.62,58.82) and (303.85,58.3) .. (304.11,58.3) .. controls (304.37,58.3) and (304.59,58.82) .. (304.83,59.37) .. controls (305.06,59.93) and (305.29,60.45) .. (305.55,60.45) .. controls (305.81,60.45) and (306.03,59.93) .. (306.27,59.37) .. controls (306.5,58.82) and (306.73,58.3) .. (306.99,58.3) .. controls (307.25,58.3) and (307.48,58.82) .. (307.71,59.37) .. controls (307.95,59.93) and (308.17,60.45) .. (308.43,60.45) .. controls (308.69,60.45) and (308.92,59.93) .. (309.15,59.37) .. controls (309.39,58.82) and (309.61,58.3) .. (309.87,58.3) .. controls (310.13,58.3) and (310.36,58.82) .. (310.59,59.37) .. controls (310.83,59.93) and (311.05,60.45) .. (311.31,60.45) .. controls (311.57,60.45) and (311.8,59.93) .. (312.03,59.37) .. controls (312.27,58.82) and (312.49,58.3) .. (312.75,58.3) .. controls (313.01,58.3) and (313.24,58.82) .. (313.47,59.37) .. controls (313.71,59.93) and (313.93,60.45) .. (314.19,60.45) .. controls (314.45,60.45) and (314.68,59.93) .. (314.91,59.37) .. controls (314.97,59.25) and (315.02,59.12) .. (315.08,59) ;
%Shape: Wave [id:dp09565149274230988] 
\draw   (348.26,49.91) .. controls (347.97,49.61) and (347.69,49.33) .. (347.44,49.43) .. controls (347.19,49.54) and (347.03,50) .. (346.87,50.49) .. controls (346.71,50.98) and (346.56,51.44) .. (346.3,51.55) .. controls (346.05,51.66) and (345.77,51.37) .. (345.48,51.08) .. controls (345.19,50.78) and (344.91,50.5) .. (344.66,50.6) .. controls (344.4,50.71) and (344.25,51.17) .. (344.09,51.66) .. controls (343.93,52.15) and (343.77,52.61) .. (343.52,52.72) .. controls (343.27,52.82) and (342.99,52.54) .. (342.69,52.24) .. controls (342.4,51.95) and (342.12,51.66) .. (341.87,51.77) .. controls (341.62,51.88) and (341.46,52.34) .. (341.3,52.83) .. controls (341.14,53.32) and (340.99,53.78) .. (340.73,53.89) .. controls (340.48,53.99) and (340.2,53.71) .. (339.91,53.41) .. controls (339.62,53.12) and (339.34,52.83) .. (339.09,52.94) .. controls (338.83,53.04) and (338.68,53.51) .. (338.52,54) .. controls (338.36,54.48) and (338.2,54.95) .. (337.95,55.06) .. controls (337.7,55.16) and (337.42,54.88) .. (337.12,54.58) .. controls (336.83,54.28) and (336.55,54) .. (336.3,54.11) .. controls (336.05,54.21) and (335.89,54.68) .. (335.73,55.17) .. controls (335.57,55.65) and (335.42,56.12) .. (335.16,56.22) .. controls (334.91,56.33) and (334.63,56.05) .. (334.34,55.75) .. controls (334.05,55.45) and (333.77,55.17) .. (333.52,55.28) .. controls (333.26,55.38) and (333.11,55.85) .. (332.95,56.33) .. controls (332.79,56.82) and (332.63,57.29) .. (332.38,57.39) .. controls (332.13,57.5) and (331.85,57.22) .. (331.55,56.92) .. controls (331.26,56.62) and (330.98,56.34) .. (330.73,56.44) .. controls (330.48,56.55) and (330.32,57.01) .. (330.16,57.5) .. controls (330,57.99) and (329.85,58.45) .. (329.59,58.56) .. controls (329.34,58.67) and (329.06,58.38) .. (328.77,58.09) .. controls (328.48,57.79) and (328.2,57.51) .. (327.95,57.61) .. controls (327.69,57.72) and (327.54,58.18) .. (327.38,58.67) .. controls (327.22,59.16) and (327.06,59.62) .. (326.81,59.73) .. controls (326.56,59.83) and (326.28,59.55) .. (325.98,59.25) .. controls (325.69,58.96) and (325.41,58.67) .. (325.16,58.78) .. controls (325.02,58.84) and (324.92,59) .. (324.82,59.21) ;
%Straight Lines [id:da8643100726237767] 
\draw    (314.53,59.18) -- (324.87,59.08) ;

% Text Node
\draw (82,49.2) node [anchor=north west][inner sep=0.75pt]    {$\hat{X}_{IJKLM}^{\ abcd} \equiv g_{ef} T_{IJK}^{eab} T_{L0M}^{cfd} =$};
% Text Node
\draw (259.17,63.4) node [anchor=north west][inner sep=0.75pt]  [font=\tiny]  {$a,J$};
% Text Node
\draw (258.84,43.4) node [anchor=north west][inner sep=0.75pt]  [font=\tiny]  {$b,K$};
% Text Node
\draw (351.51,43.4) node [anchor=north west][inner sep=0.75pt]  [font=\tiny]  {$c,L$};
% Text Node
\draw (351.84,63.07) node [anchor=north west][inner sep=0.75pt]  [font=\tiny]  {$d,M$};
% Text Node
\draw (309.67,49.83) node [anchor=north west][inner sep=0.75pt]  [font=\tiny]  {$I$};
\end{tikzpicture}
\end{eqnarray}
In summary, only five families of tensors, $\{T^2, ST, \tilde{X}, \hat{X}, X\}$, can be used to construct all the dimensionally dependent identities and we will commonly denote them as $H$, a tensor built from coloured $\nabla\F\nabla\F$. In order to consider quantities involving a pair of five antisymmetrised indices, as required to construct scalar DDIs in four dimensions, they should be completed by products of the coloured field strength tensors, which we will call the completions of $H$.

\begin{figure}
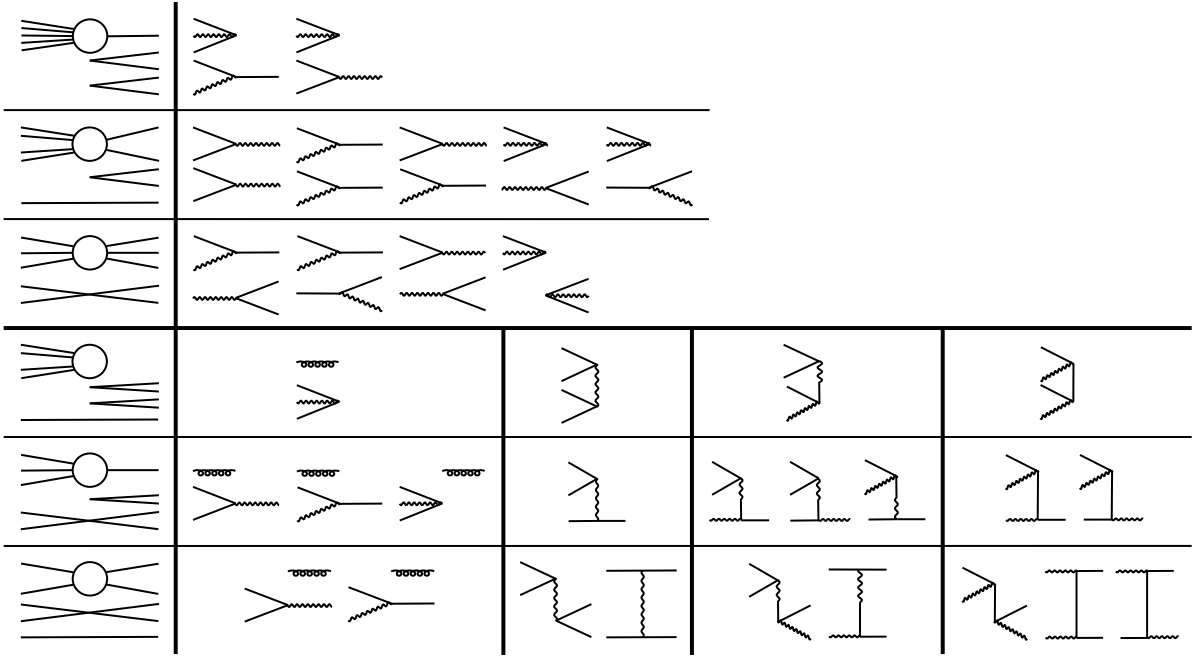

%
\caption{Classification of scalar dimensionally dependent identities  quadratic in $\nabla\F$. The first column lists the six  general types of DDIs with products of coloured field strengths, while the following columns detail the possible structures for the $H$ tensor, represented by a circle in the first column. First three rows: diagrams of the form $\{T^2\}$. Second three rows: diagrams of the forms $\{ST, \tilde{X}, \hat{X}, X\}$.}
\label{Figure1}
\end{figure}

In order to present an explicit classification of DDIs, we now wish to illustrate how they can be represented using the previous diagrams. A useful example of DDI with a coloured six-index tensor $H$ is given by
\begin{eqnarray}
\begin{tikzpicture}[x=0.75pt,y=0.75pt,yscale=-1,xscale=1]

%Straight Lines [id:da49010983133427666] 
\draw    (334.5,180.07) -- (394.81,180.13) ;
%Straight Lines [id:da054002524759643844] 
\draw    (334.5,170.07) -- (394.81,170.13) ;
%Straight Lines [id:da31230606474024436] 
\draw    (334.5,150.07) -- (394.81,150.13) ;
%Straight Lines [id:da2431499685555838] 
\draw    (334.39,140.22) -- (394.5,160.07) ;
%Straight Lines [id:da26707055640424104] 
\draw    (334.19,160.01) -- (394.5,140.07) ;

% Text Node
\draw (321.49,153.59) node [anchor=north west][inner sep=0.75pt]  [font=\tiny]  {$J_{3}$};
% Text Node
\draw (321,143.43) node [anchor=north west][inner sep=0.75pt]  [font=\tiny]  {$J_{2}$};
% Text Node
\draw (321.2,133.59) node [anchor=north west][inner sep=0.75pt]  [font=\tiny]  {$J_{1}$};
% Text Node
\draw (397.19,173.52) node [anchor=north west][inner sep=0.75pt]  [font=\tiny]  {$A_{2}$};
% Text Node
\draw (397.07,163.46) node [anchor=north west][inner sep=0.75pt]  [font=\tiny]  {$A_{1}$};
% Text Node
\draw (396.75,153.32) node [anchor=north west][inner sep=0.75pt]  [font=\tiny]  {$I_{3}$};
% Text Node
\draw (396.61,143.2) node [anchor=north west][inner sep=0.75pt]  [font=\tiny]  {$I_{2}$};
% Text Node
\draw (396.64,133.22) node [anchor=north west][inner sep=0.75pt]  [font=\tiny]  {$I_{1}$};
% Text Node
\draw (396.5,122.47) node [anchor=north west][inner sep=0.75pt]  [font=\tiny]  {$b_{i}$};
% Text Node
\draw (321.17,122.4) node [anchor=north west][inner sep=0.75pt]  [font=\tiny]  {$a_{i}$};
% Text Node
\draw (31.33,150.73) node [anchor=north west][inner sep=0.75pt]    {$\Delta_{a_{1} a_{2} a_{3} a_{4} a_{5} b_{1} b_{2} b_{3} b_{4} b_{5}} H_{J_{1} J_{2} J_{3} I_{1} I_{2} I_{3}}^{a_{1} a_{2} a_{3} b_{1} b_{2} b_{3}} F_{A_{1}}^{a_{4} b_{4}} F_{A_{2}}^{a_{5} b_{5}} \equiv $};
\end{tikzpicture}
\label{DDIexample}
\end{eqnarray}
where $\Delta$ is the partially antisymmetric tensor defined by \eqref{PAST}. A very similar example is obtained by replacing  $F_{A_{1}}^{a_{4} b_{4}} F_{A_{2}}^{a_{5} b_{5}}$ with  $F_{A_{1}}^{a_{4} a_{5}} F_{A_{2}}^{b_{4} b_{5}}$  in the above expression. In what  follows, it will be convenient to consider simultaneously these two possible configurations by using the compact representation\footnote{It is clear that if $A_i=0$ in the first configuration, the overall DDI would contain a trivial factor $(d-4)$, so that $A_i \in \{1,2,3\}$, while the antisymmetry in the second configuration imposes $A_i \in \{1,3\}$.}:
 \begin{eqnarray}
\label{Compact}
 \end{eqnarray}
The example (\ref{DDIexample}), extended with (\ref{Compact}), corresponds to the third row of Fig.\ref{Figure1}.

All the other relevant diagrams are presented in the other rows of Fig.\ref{Figure1}. 
There are two main types of DDIs, depending on whether the $H$ tensor has six (first three rows) or four indices (last three rows). The left column indicates the general form of the DDIs, with their most general completions by coloured field strengths and irrespectively of which specific tensors $H$ they are made of. The composition of the $H$ tensor, together with the specific positions of indices involved in the antisymmetrisations are filling the rest of the table. The first three rows correspond to diagrams of the form $T^2$, where all indices of both vertex tensors appear as indices of $H$, while the last three rows, with a  4-index $H$ tensor,   are divided into  the four families $\{ST, \tilde{X}, \hat{X}, X\}$, corresponding to various internal contractions of the vertex tensors.

\subsection{Colour bounds and Laplace expansions}

Although the previous DDIs are in principle straightforward to compute, there is a huge number of possibilities, or ``valency", associated with the colour indices. For instance, in the previous example \eqref{DDIexample}, a naive estimate would give $4^8$ possibilities. However, given that every (scalar) DDI can be expanded in terms of the 124 scalar invariants classified in \eqref{Agraph} and \eqref{Bgraph}, their total weight must satisfy the following constraint,
\begin{eqnarray}
0 \leq \sum_i I_i + \sum_j J_j + \sum_k A_k + K \leq 9\,, \label{GlobalBound}
\end{eqnarray}
where $I_i$ and $J_j$ are respectively the colours of $H$ associated with the indices $b_i$ and $a_i$,  $A_k$ are those involved in the completion (see the example \eqref{DDIexample}) and finally $K$ is the colour of the internal link which is present in the last three rows ($K=0$ for the other diagrams).

This global bound on colours has a local counterpart for each term of the expanded DDIs. To see this, let us consider as an example the dressed partially antisymmetric tensor of the third row of Fig.\ref{Figure1}, obtained by ``undressing" the dressed vertices and transferring their colours  to the antisymmetric tensor :
\begin{eqnarray}
{}_{A_1}^{A_2}\!\Delta^{a_1a_2a_3b_1b_2b_3}_{J_1 J_2 J_3 I_1 I_2 I_3} \equiv \Delta_{\bar{a}_1\bar{a}_2\bar{a}_3\bar{a}_4\bar{a}_5\bar{b}_1\bar{b}_2\bar{b}_3\bar{b}_4\bar{b}_5} F_{J_1}^{\bar{a}_1 a_1} F_{J_2}^{\bar{a}_2 a_2} F_{J_3}^{\bar{a}_3 a_3}  F_{I_1}^{\bar{b}_1 b_1} F_{I_2}^{\bar{b}_2 b_2} F_{I_3}^{\bar{b}_3 b_3}  F_{A_1}^{\bar{b}_4\bar{a}_4}F_{A_2}^{\bar{b}_5\bar{a}_5}\,, \label{defL3II}
\end{eqnarray}
where the ``bare"  $\Delta$ is  defined by \eqref{PAST}. 

Using the Laplace expansion, or cofactor expansion,  for determinants with respect to  $\bar{a}_5$, we get 
\bea
\label{expansion}
{}_{A_1}^{A_2}\!\Delta^{a_1a_2a_3b_1b_2b_3}_{J_1 J_2 J_3 I_1 I_2 I_3} \approx - \left( {}_{\bar{A}_1}\!\Delta^{a_1a_2a_3b_1b_2b_3}_{J_1 J_2 J_3 I_1 I_2 I_3}+{}_{A_1}\!\Delta^{a_1a_2a_3b_1b_2b_3}_{J_1 J_2 J_3 \bar{I}_1 I_2 I_3}+{}_{A_1}\!\Delta^{a_1a_2a_3b_1b_2b_3}_{J_1 J_2 J_3 I_1 \bar{I}_2 I_3}+{}_{A_1}\!\Delta^{a_1a_2a_3b_1b_2b_3}_{J_1 J_2 J_3 I_1 I_2 \bar{I}_3}\right)
\eea
with $\bar{I}_i = I_i +A_2$, $\bar{A}=A_1+A_2$ and where,  in analogy with (\ref{defL3II}), we have introduced
\beq
{}_A\Delta^{a_1a_2a_3b_1b_2b_3}_{J_1 J_2 J_3 I_1 I_2 I_3} \equiv \Delta_{\bar{a}_1\bar{a}_2\bar{a}_3\bar{a}_4\bar{b}_1\bar{b}_2\bar{b}_3\bar{b}_4} F_{J_1}^{\bar{a}_1 a_1} F_{J_2}^{\bar{a}_2 a_2} F_{J_3}^{\bar{a}_3 a_3}  F_{I_1}^{\bar{b}_1 b_1} F_{I_2}^{\bar{b}_2 b_2} F_{I_3}^{\bar{b}_3 b_3}  F_{A}^{\bar{b}_4\bar{a}_4}\,,
\eeq
the bare partially antisymmetric tensor with eight indices being defined similarly to \eqref{PAST}. The notation $\approx$ signifies that we dropped the terms with $f_2$ and $f_4$ arising from the contraction, as explained in the last section of this appendix. Notice that the relation (\ref{expansion}) enables us to relate four-dimensional DDIs, which involve antisymmetrisation over {\it five} indices,  to three-dimensional DDIs, with antisymmetrisation over {\it four} indices.

Similarly, expanding the determinant in $\bar{a}_4$ yields
\beq
{}_A\Delta^{a_1a_2a_3b_1b_2b_3}_{J_1 J_2 J_3 I_1 I_2 I_3}\approx - \left( \Delta^{a_1a_2a_3b_1b_2b_3}_{J_1 J_2 J_3 \bar{I}_1 I_2 I_3} +\Delta^{a_1a_2a_3b_1b_2b_3}_{J_1 J_2 J_3 I_1 \bar{I}_2 I_3}+\Delta^{a_1a_2a_3b_1b_2b_3}_{J_1 J_2 J_3 I_1 I_2 \bar{I}_3}\right)\,,
\eeq
wheree $\bar{I}_i = I_i +A$ ,   $(123)$ are the permutations of three elements and 
\begin{eqnarray}
\begin{split}
\Delta^{a_1a_2a_3b_1b_2b_3}_{J_1 J_2 J_3 I_1 I_2 I_3} &\equiv \Delta_{\bar{a}_1\bar{a}_2\bar{a}_3\bar{b}_1\bar{b}_2\bar{b}_3} F_{J_1}^{\bar{a}_1 a_1} F_{J_2}^{\bar{a}_2 a_2} F_{J_3}^{\bar{a}_3 a_3}  F_{I_1}^{\bar{b}_1 b_1} F_{I_2}^{\bar{b}_2 b_2} F_{I_3}^{\bar{b}_3 b_3} \\
&=\sum_{\sigma \in (123)}  (-1)^{I_1+I_2+I_3+\text{sgn}(\sigma)}   F_{I_{\sigma_1}+J_1}^{b_{\sigma_1} a_1} F_{I_{\sigma_2}+J_2}^{b_{\sigma_2} a_2}F_{I_{\sigma_3}+J_3}^{b_{\sigma_3} a_3} 
\end{split}\label{2DDDIs}
\end{eqnarray}
correspond to two-dimensional DDIs. Since $f_2\approx f_4 \approx 0$, each term in \eqref{2DDDIs}, and thus in the four-dimensional DDI \eqref{defL3II}, that contains a $F_{I_i+J_j}$ with $I_i+J_j \geq 4$ can be neglected, due to the Cayley-Hamilton identity. This results in significant valency reductions and  enables us to easily obtain specific representatives for the basis of scalar invariants.

\subsection{Algebraic basis of scalar invariants in two, three and four dimensions}

Although we are mainly interested in four-dimensional theories, it is instructive, especially after our discussion in the previous section, to consider the same theories in lower dimensions and identify the corresponding DDIs and Bianchi identities.

We observe that the two and four dimensional DDIs can all be obtained from the disconnected diagrams of the third row of Fig.\ref{Figure1}, while three-dimensional ones require the use of the sixth row as well\footnote{Furthermore, we have observed that, for each row, the two states \eqref{Compact} yield equivalent DDIs and the coloured field strengths $F_{A}$ of the completions can be set to their minimal value $A=1$. As seen in the next section, these properties are reminiscent of the various forms of the Cayley-Hamilton identity. However, notice that it might not be true when the tensor $H$ has fewer symmetries than $\nabla F\nabla F$.}. Interestingly, when the Bianchi identity is taken into account, four-dimensional DDIs can all be parametrised by the second column of the third row, corresponding to \eqref{L3IIC2} in the main text, i.e. 
\begin{eqnarray}
{}_{A_1}^{A_2}\!\Delta^{a_1a_2a_3b_1b_2b_3}_{J_1 J_2 J_3 I_1 I_2 I_3}\nabla_{a_1} \! F_{a_2 b_1} \nabla_{b_2}\!F_{b_3a_3} =   \Delta_{a_1a_2a_3a_4a_5 b_1b_2b_3b_4b_5} F_{A_1}^{b_4a_4}F_{A_2}^{b_5a_5} T^{a_1 a_2 b_1}_{J_1J_2I_1} T^{b_2 b_3 a_3}_{I_2I_3J_3}\,,
\end{eqnarray}
where the dressed partially antisymmetric tensor is defined by \eqref{defL3II}.

In two dimensions, the Bianchi identity is a tensorial DDI and any scalar is proportional to a single invariant given by 
\begin{eqnarray}
\mathscr{B}_2= \{ \B_{000}  \}\,.
\end{eqnarray} 
In three dimensions, there are 113 DDIs and only three left-over Bianchi identities, so that the algebraic basis of U(1) invariants is eight-dimensional and can be chosen to be 
\begin{eqnarray}
\mathscr{B}_3=\mathscr{B}_2 \cup  \{\A_{000},  \B_{010}, \B_{011}, \B_{110},\B_{200}, \B_{210}, \B_{211} \}\,.
\end{eqnarray}
When the U(1) symmetry is not imposed, in which case the scalars are simply (non-linear) kinetic terms for 2-form fields, the basis is eleven-dimensional and given by
\begin{eqnarray}
\tilde{\mathscr{B}}_3=\mathscr{B}_3\cup \{ \Abis_{000},\A_{010}, \A_{200} \}.
\end{eqnarray}
Finally, in four dimensions, we obtain 22 U(1) invariants, which can be chosen as follows,
\begin{eqnarray}
\mathscr{B}_4=\mathscr{B}_3\cup \{ \A_{002}, \A_{101}, \A_{200}, \A_{210}, \A_{301}, \B_{020}, \B_{022}, \B_{112}, \B_{120}, \B_{212}, \B_{220}, \B_{310}, \B_{312}, \B_{320} \}
\end{eqnarray}
while we obtain 36 independent invariants for the kinetic terms of 2-form fields,
\begin{eqnarray}
\tilde{\mathscr{B}}_4=\mathscr{B}_4 \cup \{\A_{010}, \A_{021}, \A_{120}, \A_{320}, \Abis_{000}, \Abis_{101}, \Abis_{110}, \Abis_{200}, \Abis_{301}, \Abis_{310}, \Bbis_{011}, \Bbis_{211}, \Bter_{001}, \Bter_{201}\}
\end{eqnarray}

\subsection{Redundancy of 2-index $H$ tensor}

As claimed above, we show that if 2-index tensors, such as $\V\U$ or one of the loop-shaped tensors (see the diagram \eqref{DefTtensor}),
\begin{eqnarray}
g_{be}\, g_{cf} T^{abc}_{000}T^{def}_{IJK} \,,
\end{eqnarray}
are used to construct DDIs, the result will end up being redundant compare to the Cayley-Hamilton  (CH) identity. In order to do so, notice that this identity can be obtained from both
\begin{eqnarray}
\delta^a_{[b} F^{e_1}{}_{e_1} F^{e_2}{}_{e_2} F^{e_3}{}_{e_3} F^{e_4}{}_{e_4]} \propto \delta^a_{[b} F_{a_1 a_2}F_{a_3 a_4]}F^{a_1 a_2}F^{a_3 a_4} =0\,.\label{CHDDI}
\end{eqnarray}
Noticing further that
\begin{eqnarray}
\delta^a_{[b} F_{a_1 a_2}F_{a_3}{}^{a_3}F_{a_4]}{}^{a_4}F^{a_1 a_2}=0\,,
\end{eqnarray} 
in any dimension, this shows that DDIs with two indices constructed solely from products of the field strength are either trivial or redundant compared to CH, given that any higher weight can be obtained multiplying the previous equations by $F_I$ and that we took into consideration the three different ways to antisymmetrise five indices from products of $F$. Notice that colouring each $F$ in the previous equations does not change that result either.

Denoting $H$ an arbitrary 2-index tensor (which for our purpose should be constructed from $\nabla\F\nabla\F$), the DDIs built from it are divided into two families,
\begin{eqnarray}
\begin{split}
H_{[a_1 a_2}\left( F_I F_J F_K F_L\right)_{a_3 a_4 a_5]}^{a_1 a_2 a_3 a_4 a_5} &= H_{a_1 a_2}\left( F_I F_J F_K F_L\right)_{a_3 a_4 a_5}^{[a_1 a_2 a_3 a_4 a_5]}, \\
H_{[a_1}^{a_1}\left( F_I F_J F_K F_L\right)_{a_2 a_3 a_4 a_5]}^{a_2 a_3 a_4 a_5} &= H_{b_1}^{a_1} \delta_{[a_1}^{b_1}\left( F_I F_J F_K F_L\right)_{a_2 a_3 a_4 a_5]}^{a_2 a_3 a_4 a_5},
\end{split}
\end{eqnarray} 
where the notation with parenthesis means that the indices can be distributed arbitrarily on the tensor product. Thus, they reduce to tensorial DDIs with two indices built from products of $F_I$, i.e. to the Cayley-Hamilton identity. 

\subsection{Connected scalar diagrams}
 
Finally, we explain why it is possible to neglect the traces $f_2\approx f_4 \approx 0$ in the expansion of generalised Kronecker symbols, without spoiling the dimensionality of the basis of DDIs. Although not necessary, this assumption helps to find representatives for that basis. Of course, in order to obtain precise irreducible decompositions of invariants, this simplification cannot be made. In terms of diagrams, this means that among all the scalar invariants resulting from Laplace expansion, only those represented by connected diagrams can be kept. More precisely, it is clear that any scalar DDI can be expanded to yield
\begin{eqnarray}
\mathcal{S}_w = \sum_{2\leq 2(i+2j)\leq w \leq 9} f_2^{i}\, f_4^{j} \, \mathcal{S}_{w-2(i+2j)}^{(i,j)}\,, \label{DDIexpansionf2f4}
\end{eqnarray}
where $\mathcal{S}_w$ is a linear combination with constant coefficients of the scalar invariants $\{\A, \Abis, \B, \Bbis, \Bter \}$ of weight $w$, defined by \eqref{Agraph} and \eqref{Bgraph}, and similarly for each $\mathcal{S}^{(i,j)}$. In order to list all independent DDIs, we now show that it is sufficient to  set $f_2\approx f_4 \approx 0$.

Indeed,  this simplification would lead to ignore some relevant DDIs only if  two DDIs yield \eqref{DDIexpansionf2f4} with the same $\mathcal{S}_w$ but different $\mathcal{S}^{(i,j)}$. Denoting  $\widetilde{\mathcal{S}}$ this difference of invariants, we could then construct  a DDI given by 
\begin{eqnarray}
 \sum_{2\leq 2(i+2j)\leq w \leq 9} f_2^{i} f_4^{j}  \widetilde{\mathcal{S}}_{w-2(i+2j)}^{(i,j)}  =0, \label{limpossible}
\end{eqnarray}
which would trivialize when assuming $f_2\approx f_4 \approx 0$. However, the unicity of the Laplace expansions \eqref{DDIexpansionf2f4} prevents this from happening.

Let us see some examples for low weights, which we recall, are ranging from 0 to 9. First, it is clear that at least 10 indices are needed to build a four-dimensional scalar DDI, so $w \geq 2$ in \eqref{DDIexpansionf2f4}. At weight 2, all DDIs have the form 
\begin{eqnarray}
\mathcal{S}_2 = f_2 \mathcal{S}_0^{(1,0)} \,.\label{Bonjour}
\end{eqnarray}
Thus, \eqref{limpossible} would imply the existence of a dimensional identity between weight 0 scalars, $\widetilde{\mathcal{S}}_0=0$, which is impossible as we just said. The same would happen at weight 3, so we move to $w=4$. At this weight, DDIs have the form 
\begin{eqnarray}
\mathcal{S}_4 = f_2 \mathcal{S}_2^{(1,0)}+ f_2^2 \mathcal{S}_0^{(2,0)}+ f_4 \mathcal{S}_0^{(0,1)}\,.
\end{eqnarray}
Thus, a relation of the form \eqref{limpossible} would yield $f_2 \widetilde{\mathcal{S}}_2^{(1,0)}+ f_2^2 \widetilde{\mathcal{S}}_0^{(2,0)}+ f_4 \widetilde{\mathcal{S}}_0^{(0,1)} =0$. Assuming that we have used all the lower weight DDIs, given by \eqref{Bonjour}, implies that $\widetilde{\mathcal{S}}_0^{(0,1)}\neq 0$. Let us now choose a configuration of the two-form field $F$ which fixes one out of its six components, say $f_2=0$ and $f_4\neq 0$. Then, the previous equation becomes a DDI among weight 0 scalar invariants, which is again impossible, unless $F$ is effectively two-dimensional in this configuration, meaning that it would have only one independent component. This is not the case here as only one component among six has been fixed, so again relations such as \eqref{limpossible} are impossible up to weight 4. It is clear that the same would apply for higher weight terms.

\bibliographystyle{utphys}
\bibliography{Biblio_DEM_classification_arxiv2}
  
\end{document}